\documentclass[11pt]{article}

\usepackage[margin=1in]{geometry}
\usepackage[T1]{fontenc}
\usepackage{lmodern}
\usepackage[numbers,sort&compress]{natbib}
\usepackage{amsmath}
\usepackage{amssymb}
\usepackage{booktabs}
\usepackage{graphicx}
\usepackage{tabularx}
\usepackage[table]{xcolor}
\usepackage{pgfplots}
\usepackage{tikz}
\usepackage{placeins}
\usepgfplotslibrary{groupplots}
\usepgfplotslibrary{fillbetween}
\pgfplotsset{compat=1.18}
\usepackage{authblk}
\usepackage[hidelinks]{hyperref}
\hypersetup{
    pdftitle={Empirical Calibration and Conditional-Reliability Diagnostics for Bearing RUL Prediction under Operating-Regime Shift},
    pdfauthor={Shaoliang Yang, Jun Wang, Yunsheng Wang},
    pdfkeywords={reliability, prognostics, remaining useful life, calibration, conformal prediction, operating-regime shift, bearing}
}

\title{Empirical Calibration and Conditional-Reliability Diagnostics for Bearing RUL Prediction under Operating-Regime Shift}
\author[1]{Shaoliang Yang}
\author[1]{Jun Wang\thanks{Corresponding author. E-mail: \texttt{jwang22@scu.edu}}}
\author[1]{Yunsheng Wang}
\affil[1]{Department of Mechanical Engineering, Santa Clara University, Santa Clara, CA 95053, USA}
\date{July 9, 2026}
\newcommand{\publiccodereleaseurl}{\href{https://github.com/nbbllxx0/Empirical-Calibration-and-Conditional-Reliability-Diagnostics}{\texttt{https://github.com/\allowbreak nbbllxx0/\allowbreak Empirical-\allowbreak Calibration-\allowbreak and-\allowbreak Conditional-\allowbreak Reliability-\allowbreak Diagnostics}}}

\begin{document}

\maketitle

\begin{abstract}
Remaining useful life (RUL) estimates support reliability and maintenance decisions only if both point accuracy and prediction intervals remain trustworthy when operating conditions change. Convenient mixed splits can hide that failure. This paper studies the question on a documented 10-bearing PHME subset with time-varying load and speed. Derived load--speed regimes define the held-out evaluation units, while models receive only measured load and speed as context. A calibrated predictive-representation model fuses raw vibration windows, engineered descriptors, and operating context, then forms intervals by empirical residual calibration. Under strict train/validation/calibration/test separation, the model reaches normalized MAE 0.1477, empirical 90\% coverage 0.900, and retrospective absolute-step MAE 285.26; a 400-tree random forest reaches 0.1538, 0.871, and 294.57. The results do not show uniform dominance: conditional diagnostics expose non-uniform reliability, including 0.666 coverage in a low-load/high-speed cell, and a post-hoc pooled regime-conditioned residual diagnostic raises that cell to 0.941 only as motivation for future pre-specified conditional calibration. Stress tests further identify raw-channel loss as the largest tested reliability failure mode. The contribution is therefore a bounded reliability-evaluation protocol for the processed 10-bearing subset, with conditional undercoverage and raw-channel loss reported explicitly as failure modes rather than deployment guarantees.
\end{abstract}

\noindent\textbf{Keywords:} reliability; prognostics; remaining useful life; calibration; conformal prediction; operating-regime shift; bearing

\section{Introduction}

Rolling bearings rarely fail under a single fixed load and speed. In service they see changing duty cycles, and those changes reshape both vibration signatures and degradation rates. Remaining useful life (RUL) prediction is meant to turn condition-monitoring signals into time-to-failure estimates that can support inspection and replacement decisions \citep{jardine2006cbm,lei2018machinery_review,zio2022phm}. For bearings, public accelerated-degradation experiments have made vibration-based RUL a mature research setting \citep{nectoux2012pronostia,wang2020hybrid_bearing}. The open question is no longer whether a model can fit a convenient split; it is whether accuracy and uncertainty remain trustworthy when operating conditions shift.

That shift is the practical bottleneck. A model trained under one load--speed mixture can degrade under another because the signal distribution and the remaining-life mapping both move. Reliability modeling has treated time-varying operating conditions with state-space degradation formulations \citep{li2019twofactor_tvoc}, while data-driven work has pursued cross-condition transfer and domain adaptation \citep{dacosta2020domain_adaptation,zhang2021representation_regularization,ding2021deep_metric_transfer}. The PHME time-varying operating-condition bearing archive makes the issue concrete: run-to-failure experiments with measured load and speed variation \citep{javanmardi2024phme_tvoc,aimiyekagbon2024zenodo}.

Methodologically, the surrounding literature is already dense. Deep encoders, recurrent and convolutional predictors, graph fusion models, and PHM surveys establish learned RUL representations as standard tools rather than novelty claims \citep{li2018dcnn_rul,zhang2018lstm_rul,li2019multiscale_bearing,wang2019dscn,li2021hagcn,fink2020deep_phm}. Digital-twin and physics-informed studies cover much of the language that can otherwise sound new \citep{cui2024dt_graph_da,lu2023dt_bearing,gong2025dt_pinn_uq,liao2023attnpinn,li2024pidd_rul_review}. This paper therefore does not present a new digital twin, a new fatigue theory, or a full-PHME leaderboard result.

The gap we address is narrower and more reliability-facing. Under time-varying bearing regimes, point error alone is not enough. Useful evaluation should ask whether a model still works when the test regime is held out, whether nominal intervals actually cover, whether coverage is uniform across regimes and bearings, how early-life prefixes behave, and what happens under sensing stress. A method can look strong under mixed splits and still fail in a low-load/high-speed cell, under raw-channel loss, or as a maintenance trigger. Uncertainty statements are decision-relevant only when those failure modes are measured rather than assumed away \citep{sankararaman2015uncertainty,javanmardi2023conformal_rul}. We therefore prioritize a transparent, bounded evaluation protocol over full-archive breadth.

Our approach follows that evaluation focus. We study a documented 10-bearing PHME subset under leave-operating-regime-out evaluation. Derived load--speed regimes define the split and reporting units, while models receive only measured load and speed as context. A calibrated predictive-representation model fuses raw vibration windows, engineered descriptors, and operating context, then forms intervals by empirical residual calibration. The representation is a predictive intermediate, not a validated physical damage coordinate. Weak monotonicity or smoothness regularizers appear only as optional diagnostics, not as the claimed source of gains.

The paper makes five contributions:
\begin{enumerate}
    \item A regime-disjoint reliability protocol for bearing RUL under time-varying load and speed on a documented PHME subset.
    \item A strict calibration design that separates model selection, residual calibration, and held-out regime testing.
    \item A predictive representation that combines raw vibration, engineered features, and measured operating context, evaluated as a reliability baseline rather than a physics twin.
    \item Conditional reliability diagnostics across regimes, bearings, prefix windows, and stress settings, including explicit undercoverage cells.
    \item Retrospective maintenance-risk diagnostics at the bearing-regime unit, without claiming a prospective alarm policy.
\end{enumerate}

On the 10-bearing analysis set, the primary endpoint is the strict four-way protocol with separate train, validation, calibration, and test roles. The calibrated model reaches a normalized MAE of 0.1477 with empirical 90\% coverage of 0.900. A 400-tree random forest is close, with normalized MAE 0.1538 and coverage 0.871. The margin is modest and subset-specific: the model is not uniformly point-error dominant. Its practical value is the accuracy--coverage tradeoff under regime-disjoint calibration. Stress tests further identify raw-channel loss as the largest tested reliability failure mode, and the unprocessed remainder of the public PHME archive remains outside the evidence boundary.

\section{Related Work}

Bearing RUL has been studied extensively across diagnostics, prognostics, and deep learning. Rather than catalog every architecture, this section organizes prior work by the questions it addresses---the problem formulation, the evidence used, and the open limitations---and then situates the present study relative to those answers.

\subsection{From Condition Monitoring to RUL Decision Support}

RUL estimation is one step in a broader reliability workflow. Jardine et al.\ review condition-based maintenance and treat diagnostics and prognostics as routes from monitoring data to maintenance action \citep{jardine2006cbm}. Heng et al.\ survey rotating-machinery prognostics and stress that laboratory success does not automatically transfer to variable field conditions \citep{heng2009rotating}. Sikorska et al.\ compare industrial prognostic options and show that the right model depends on data quality, failure definition, and the decision to be supported \citep{sikorska2011prognostic}. Si et al.\ organize statistical data-driven RUL methods by how degradation and remaining life are represented under uncertainty \citep{si2011statistical_rul}. Lei et al.\ provide a machinery-health-prognostics map from data acquisition through RUL prediction \citep{lei2018machinery_review}. Later PHM reviews by Zio and by Hu et al.\ restate the field as a design--development--decision loop rather than a single forecasting recipe \citep{zio2022phm,hu2022phm_review}. Sankararaman argues that prognostic uncertainty must be interpretable if intervals are to guide decisions rather than decorate point forecasts \citep{sankararaman2015uncertainty}.

Taken together, these reviews set the standard of judgment used here: a useful RUL method is one whose accuracy and uncertainty statements remain meaningful under the operating conditions where decisions would be made.

\subsection{Bearing Benchmarks and Hybrid Prognostics}

Public bearing experiments made vibration RUL a shared testbed. Nectoux et al.\ introduced PRONOSTIA for accelerated bearing degradation tests \citep{nectoux2012pronostia}. The NASA IMS archive remains a common public source for bearing monitoring studies \citep{nasa2025ims}. Wang et al.\ combined health-indicator construction with remaining-life estimation in a hybrid bearing prognostics framework and helped fix vibration RUL as a mature evaluation setting \citep{wang2020hybrid_bearing}.

These resources explain why point-prediction methods can look mature while regime-shift reliability questions remain incompletely answered: many studies optimize for established constant-condition or mixed-split benchmarks rather than for held-out load--speed cells with interval diagnostics.

\subsection{Deep Encoders for Vibration and Multi-Sensor RUL}

Deep learning then became the default representation route. Guo et al.\ built recurrent health indicators for bearing RUL \citep{guo2017rnn_hi}. Li et al.\ estimated RUL with deep convolutional networks on prognostics benchmarks \citep{li2018dcnn_rul}, while Zhang et al.\ used LSTM networks for remaining-life prediction \citep{zhang2018lstm_rul}. Ren et al.\ applied deep convolution specifically to bearings \citep{ren2018dcnn_bearing}. Li et al.\ introduced multi-scale feature extraction for bearing RUL \citep{li2019multiscale_bearing}, and Wang et al.\ proposed deep separable convolutions for machinery RUL \citep{wang2019dscn}. Zhao et al.\ and Fink et al.\ surveyed deep learning for machine health monitoring and PHM more broadly \citep{zhao2019deep_health,fink2020deep_phm}. Graph-based fusion followed: Li et al.\ used hierarchical attention graph convolutions for multi-sensor RUL \citep{li2021hagcn} and later provided a GNN guideline and benchmark for fault diagnostics and prognostics \citep{li2022gnn_benchmark}. Bai et al.\ evaluated temporal convolutional networks for sequence modeling and thereby supplied a standard source for TCN-style baselines \citep{bai2018tcn}.

These contributions establish that recurrent, convolutional, multi-scale, and graph encoders are already available tools. They do not, by themselves, certify that prediction intervals remain reliable when the held-out unit is an operating regime.

\subsection{Operating-Condition Shift and Cross-Condition Transfer}

Condition shift changes both signals and degradation rates. Li et al.\ addressed time-varying operating conditions with a two-factor state-space model that couples degradation dynamics to operating factors \citep{li2019twofactor_tvoc}. In deep transfer settings, da Costa et al.\ studied RUL prediction through deep domain adaptation across operating conditions \citep{dacosta2020domain_adaptation}. Zhang et al.\ used representation regularization for cross-condition RUL transfer \citep{zhang2021representation_regularization}. Ding et al.\ combined deep metric transfer learning with kernel regression under cross-condition mismatch \citep{ding2021deep_metric_transfer}.

This line establishes that mismatch is both a modeling problem and an evaluation problem. Most of it optimizes transfer of point predictions or domain-invariant features. Fewer studies fix a measured load--speed archive, separate residual calibration from model selection, and then report conditional coverage and sensing-stress failures under the same protocol.

\subsection{The PHME Time-Varying Load and Speed Setting}

The PHME campaign makes time-varying load and speed an explicit experimental design rather than a residual nuisance. Javanmardi et al.\ introduced RUL estimation for bearings operating under time-varying conditions and linked operating-condition-aware prediction to that campaign \citep{javanmardi2024phme_tvoc}. Aimiyekagbon released the public run-to-failure archive of ball bearings under time-varying load and speed; the deposit used here is the full 17-run Zenodo record \citep{aimiyekagbon2024zenodo}.

We use that source but do not claim full-archive PHME performance. The evidence boundary is a documented 10-bearing processed subset under leave-operating-regime-out evaluation. Derived regime labels define held-out cells; the primary model sees only measured load and speed.

\subsection{Recent Reliability-Journal Bearing RUL and Interval Work}

Recent reliability-journal papers respond to variable conditions and uncertainty with new architectures. Hou et al.\ predict bearing remaining life under variable conditions using segmented data cleaning and Cross-Transformer fusion, emphasizing prediction and generalization rather than split-controlled residual calibration \citep{hou2024cross_transformer_segmented_cleaning}. Wang et al.\ estimate rolling-element-bearing intervals with a multi-task mixture density network, producing probabilistic intervals from mixture outputs \citep{wang2024mtl_mdn_bearing_interval}. Zhan et al.\ quantify RUL uncertainty by multi-distribution fusion, learning aleatoric uncertainty through multiple output distributions \citep{zhan2024mdf_uq}. Li et al.\ pursue reliable bearing RUL through multi-hierarchy dynamic evaluation and uncertainty amelioration, focusing on adaptive degradation-state tracking \citep{li2025reliable_bearing_mhde}.

These studies are the closest recent neighbors in spirit. They mainly advance predictors or uncertainty heads. The present study instead holds the evaluation protocol fixed on measured PHME regimes and asks what empirical coverage, conditional undercoverage, stress fragility, and maintenance-trigger tradeoffs look like under that protocol.

\subsection{Digital Twins and Physics-Informed Prognostics}

Digital-twin and physics-informed ideas are already active. Tao et al.\ and Fuller et al.\ review industrial digital twins and their open challenges \citep{tao2019digital_twin,fuller2020digital_twin}. Lu et al.\ study digital-twin-oriented bearing RUL settings \citep{lu2023dt_bearing}. Cui et al.\ combine a dynamic digital twin with graph domain adaptation for rolling-bearing RUL, emphasizing twin-generated data and domain adaptation \citep{cui2024dt_graph_da}. Gong et al.\ estimate bearing RUL with digital-twin and physics-informed components plus uncertainty quantification \citep{gong2025dt_pinn_uq}. Bott et al.\ pursue uncertainty-aware ball-bearing prognostics through physics-based simulation and conditional normalizing flows \citep{bott2026uncertainty_bearing_flows}.

On the physics-informed learning side, Raissi et al.\ introduced PINNs for PDE forward and inverse problems \citep{raissi2019pinn}, and Karniadakis et al.\ survey physics-informed machine learning \citep{karniadakis2021piml}. Li et al.\ review physics-informed data-driven RUL prediction and its open challenges \citep{li2024pidd_rul_review}. Liao et al., He et al., and Beaulieu et al.\ explore attention-based, graph-based, and physics-augmented prognostics formulations \citep{liao2023attnpinn,he2023pignn,beaulieu2024physics_aug}.

These lines bound novelty. We do not introduce a high-fidelity bearing twin or a new fatigue theory, and we do not treat weak smoothness or monotonicity regularizers as proven accuracy mechanisms.

\subsection{Interval Construction and Conformal RUL}

If intervals are to support reliability decisions, their coverage under the intended protocol must be measured. Lei et al.\ develop distribution-free predictive inference under exchangeability \citep{lei2018distribution_free}. Romano et al.\ introduce conformalized quantile regression as a practical interval construction tool \citep{romano2019cqr}. Angelopoulos and Bates provide a general introduction to conformal prediction and distribution-free uncertainty quantification \citep{angelopoulos2021gentle}. Javanmardi and H{\"u}llermeier study conformal intervals for RUL estimation, wrapping RUL predictors with formal conformal framing on C-MAPSS-style settings \citep{javanmardi2023conformal_rul}.

Our calibration design is related but narrower. We use residual calibration in the spirit of split conformal practice, report empirical coverage under PHME leave-regime-out evaluation and stress, and treat exchangeability as something to inspect when regimes and sensor streams are held out. A pooled regime-conditioned residual diagnostic appears only as a post-hoc motivation for future pre-specified conditional calibration.

\subsection{Where the Present Study Sits}

Three patterns emerge. Deep and tabular predictors already compete on point error. Operating-condition shift is widely recognized, yet evaluation often remains architecture-centric. Uncertainty-aware bearing work increasingly reports probabilistic or conformal intervals, but fewer studies fix a measured time-varying subset, separate residual calibration from model selection, expose conditional undercoverage and sensing-stress failures, and keep physics claims deliberately limited.

Against that background, this paper is a bounded reliability-evaluation study: regime-disjoint testing on a documented 10-bearing PHME subset, load/speed-only context control, strong non-latent baselines including a 400-tree random forest, empirical residual calibration with conditional diagnostics, and explicit stress-test limitations.

\section{Data and Evaluation Protocol}

\subsection{Dataset Scope and Evidence Boundary}

We study the PHME time-varying operating-condition bearing archive, which provides run-to-failure vibration experiments under changing load and speed \citep{javanmardi2024phme_tvoc,aimiyekagbon2024zenodo}. The public deposit used here is the full 17-run Zenodo record 10868257 (version 2024-04-02); the PHME conference paper also cites an earlier DOI, 10.5281/zenodo.10805042. The analysis set is smaller by design. After the documented parsing and quality-control pipeline, the processed subset was fixed as B01, B02, B03, B04, B05, B08, B10, B11, B12, and B17 before final model comparison. The remaining seven public records, B06, B07, B09, B13, B14, B15, and B16, were not processed, and they total 126.96 GB in the Zenodo listing. All conclusions are therefore limited to this transparent 10-bearing subset: the study evaluates regime shift and calibration behavior there, rather than claiming full 17-bearing PHME performance or transfer to XJTU-SY, IMS, FEMTO, or C-MAPSS.

\begin{table}[!htbp]
\centering
\caption{PHME analysis set and preprocessing scope used in this study.}
\label{tab:dataset_summary}
\begin{tabular}{ll}
\toprule
Item & Value \\
\midrule
Dataset scope & 10-bearing PHME time-varying operating-condition analysis set \\
Evaluated bearing runs & B01, B02, B03, B04, B05, B08, B10, B11, B12, B17 \\
Processed windows & 14,297 \\
Derived operating regimes & 9 \\
Vibration channels used & 1 \\
Sampling rate & 25600 Hz \\
Window length & 512 samples \\
Engineered feature columns & 36 stored; 18 unique single-channel descriptors \\
\bottomrule
\end{tabular}

\vspace{0.25em}
{\footnotesize The study is limited to this 10-bearing analysis set; it is not a full 17-bearing PHME evaluation.}
\end{table}

\begin{figure}[!htbp]
\centering
\begin{tikzpicture}[x=1mm,y=1mm,every node/.style={font=\scriptsize}]
\draw[rounded corners=1pt,draw=gray!65,fill=gray!5] (0,47) rectangle (72,78);
\node[font=\small\bfseries,anchor=west] at (3,74) {(a) PHME evidence boundary};
\draw[draw=gray!55,fill=gray!12] (5,54) rectangle (31,66);
\draw[draw=blue!65!black,fill=blue!15] (39,54) rectangle (67,66);
\node at (18,60) {full public record};
\node at (53,60) {10-bearing subset};
\draw[->,thick,draw=gray!65] (31,60) -- (39,60);
\node[anchor=west] at (5,51) {analysis set fixed before evaluation};

\draw[rounded corners=1pt,draw=gray!65,fill=gray!5] (83,47) rectangle (155,78);
\node[font=\small\bfseries,anchor=west] at (86,74) {(b) Load-speed regime grid};
\foreach \x in {96,114,132,150} {\draw[gray!60] (\x,52) -- (\x,69);}
\foreach \y in {52,57.7,63.3,69} {\draw[gray!60] (96,\y) -- (150,\y);}
\node[font=\tiny] at (105,70.8) {low S};
\node[font=\tiny] at (123,70.8) {mid S};
\node[font=\tiny] at (141,70.8) {high S};
\node[font=\tiny,anchor=east] at (94,54.8) {low L};
\node[font=\tiny,anchor=east] at (94,60.5) {mid L};
\node[font=\tiny,anchor=east] at (94,66.2) {high L};
\fill[blue!18] (114,57.7) rectangle (132,63.3);
\node[font=\tiny] at (123,60.5) {test cell};

\draw[rounded corners=1pt,draw=gray!65,fill=gray!5] (0,5) rectangle (72,38);
\node[font=\small\bfseries,anchor=west] at (3,34) {(c) Split policy};
\node[draw,fill=teal!13,minimum width=13mm,minimum height=7mm] (train) at (9,24) {train};
\node[draw,fill=orange!18,minimum width=17mm,minimum height=7mm] (select) at (27,24) {\shortstack{model\\selection}};
\node[draw,fill=yellow!22,minimum width=17mm,minimum height=7mm] (cal) at (47,24) {calibration};
\node[draw,fill=red!13,minimum width=13mm,minimum height=7mm] (test) at (64,24) {test};
\draw[->,thick] (train) -- (select);
\draw[->,thick] (select) -- (cal);
\draw[->,thick] (cal) -- (test);
\node[font=\tiny,align=center,text width=60mm] at (36,13) {matched sensitivity: merged selection/calibration};

\draw[rounded corners=1pt,draw=gray!65,fill=gray!5] (83,5) rectangle (155,38);
\node[font=\small\bfseries,anchor=west] at (86,34) {(d) Reliability outputs};
\node[draw,fill=blue!12,minimum width=24mm,minimum height=7mm] at (99,24) {MAE/RMSE};
\node[draw,fill=green!14,minimum width=24mm,minimum height=7mm] at (128,24) {coverage};
\node[draw,fill=purple!12,minimum width=24mm,minimum height=7mm] at (99,13) {stress tests};
\node[draw,fill=gray!12,minimum width=24mm,minimum height=7mm] at (128,13) {leave-bearing};
\end{tikzpicture}
\caption{Four-part protocol schematic for the bounded PHME subset, derived load-speed regimes, strict primary train/validation/calibration/test split, matched-protocol sensitivity analysis, and reliability-oriented outputs.}
\label{fig:protocol_overview}
\end{figure}
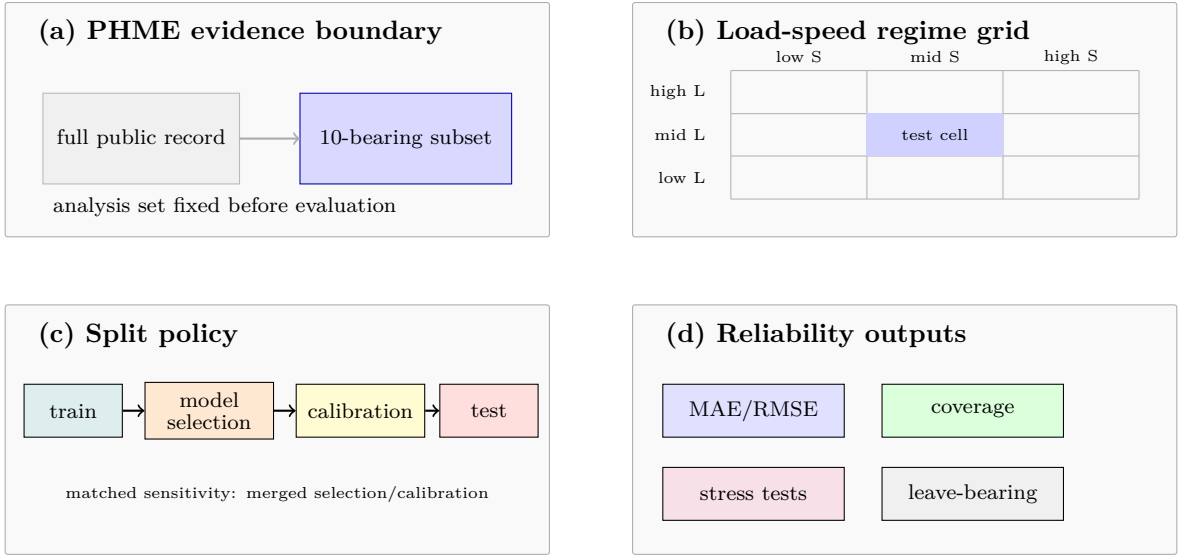

Figure~\ref{fig:protocol_overview} summarizes the evidence boundary, derived load-speed regime grid, strict primary split, and reliability-oriented outputs used in this study.

The analysis set contains 14,297 processed windows after segmentation and feature construction. Each window is treated as a supervised RUL example with bearing identity, operating context, life fraction, absolute RUL, and normalized RUL fields. Windows are nested within only ten run-to-failure trajectories and are not treated as independent inferential units; window-level metrics are descriptive, while paired statistical comparisons use held-out regimes or bearings as the paired units. The RUL label for a window is computed from the remaining step count within that bearing trajectory. This label construction avoids using windows from other bearings to define a test trajectory's endpoint, but it still assumes that the run-to-failure trajectory for the evaluated bearing is available for retrospective label construction and metric computation. That assumption is standard for offline RUL studies, but it should not be confused with online deployment where the final failure time is unknown.

\subsection{Windowing and Feature Construction}

The preprocessing pipeline creates fixed-length vibration windows of 512 samples at a nominal sampling rate of 25.6 kHz. The current processed subset uses one vibration channel. Each window is zero-padded or truncated as needed and standardized per channel by its own mean and standard deviation before feature extraction. For each normalized vibration window, the pipeline stores the normalized signal tensor and a deterministic engineered-feature row computed from that normalized window. The engineered features include time-domain statistics, impulsiveness measures, spectral summaries, band-energy measures, and envelope summaries. Because they are computed after per-window standardization, they emphasize shape, spectral distribution, and impulsiveness rather than absolute vibration amplitude. This avoids train/test normalization leakage but can suppress amplitude-degradation cues; no raw-amplitude feature ablation is claimed in this manuscript. Because the processed subset is single-channel, the stored fused descriptors duplicate the corresponding channel-0 descriptors; the manuscript reports both the stored-column count and the effective single-channel descriptor count. The primary model input uses only measured load and speed as operating context. Derived operating-regime labels are used for split construction, reporting, and sensitivity analysis, not as primary model inputs.

\begin{table}[!htbp]
\centering
\caption{Signal, feature, context, and label fields generated by preprocessing.}
\label{tab:feature_groups}
\scriptsize
\begin{tabularx}{\textwidth}{>{\raggedright\arraybackslash}p{0.19\textwidth}>{\raggedright\arraybackslash}p{0.16\textwidth}>{\raggedright\arraybackslash}p{0.15\textwidth}>{\raggedright\arraybackslash}X}
\toprule
Input group & Count/field & Role & Notes \\
\midrule
Raw vibration & 512 samples per window & Model input & Primary signal stream used by sequence models. \\
Time-domain feature columns & 18 stored columns & Model input & Stored channel-0 and fused time-domain descriptors; fused values duplicate channel-0 descriptors in this single-channel subset. \\
Spectral/envelope feature columns & 18 stored columns & Model input & Stored channel-0 and fused spectral, band-energy, and envelope descriptors; fused values duplicate channel-0 descriptors in this single-channel subset. \\
Operating context & load, speed & Model input & Primary context input. Derived regime labels are used for split construction and reporting, not as primary model input. \\
RUL / normalized RUL & RUL, normalized RUL & Supervised target & Targets are derived within each bearing trajectory after preprocessing. \\
Life fraction & life fraction & Evaluation variable & Used for stratification and diagnostics; not provided as an inference-time input. \\
\bottomrule
\end{tabularx}

\vspace{0.25em}
{\footnotesize Because the processed subset uses one vibration channel, fused descriptors duplicate the corresponding channel-0 descriptors; the table reports stored columns and distinguishes them from unique descriptor families.}
\end{table}

\begin{table}[!htbp]
\centering
\caption{Preprocessing audit for the 10-bearing PHME analysis set.}
\label{tab:preprocessing_audit}
\small
\begin{tabularx}{\textwidth}{p{0.24\textwidth}p{0.35\textwidth}X}
\toprule
Design item & Current implementation & Interpretation for this study \\
\midrule
Signal window & 512 samples from one vibration channel; windows are zero-padded or truncated as needed. & Fixed-window design used for the reported analysis; no window-length optimality claim is made. \\
Window normalization & Each window is standardized by its own channel mean and standard deviation before storage and feature extraction. & Prevents train/test scaling leakage, but suppresses absolute-amplitude cues that may carry degradation information. \\
Engineered features & Time, impulsiveness, spectral, band-energy, and envelope descriptors are computed from the standardized window. & Features represent normalized waveform shape and spectral distribution, not raw vibration amplitude. \\
Feature-column scaling & Tabular feature columns are scaled inside the model-training pipeline using training data for the corresponding split. & Evaluation should be read as split-controlled feature scaling, not full-table target leakage. \\
Operating context & Primary model receives continuous measured load and speed only. & Derived regime labels are reserved for split construction and reporting. \\
Channel policy & The processed subset uses one vibration channel; fused descriptors duplicate channel-0 descriptors in this single-channel setting. & Multi-channel benefit and channel-selection sensitivity are untested. \\
\bottomrule
\end{tabularx}
\end{table}

Figure~\ref{fig:load_speed_trajectories} shows the measured load-speed trajectories of the ten evaluated bearings, which motivate treating derived load-speed cells as the primary held-out evaluation units.

\input{figures/fig_load_speed_trajectories}

The feature design is intentionally conventional. It is not presented as a new signal-processing contribution. Its role is to provide a strong and interpretable input stream for tabular baselines and for the fused predictive-representation model. This also makes the ablation ``no engineered features'' meaningful: the model must demonstrate whether raw vibration alone is sufficient under regime shift or whether conventional descriptors still add information.

\subsection{Operating-Regime Definition}

The primary evaluation unit is a derived load-speed regime. Let $u_i=(\ell_i,s_i)$ denote the load and speed metadata for window $i$. The preprocessing code divides load and speed into empirical low, middle, and high bins using the one-third and two-third quantiles of the analysis set. Each window is then assigned one of nine low/middle/high load-speed cells, such as low-load/low-speed, mid-load/high-speed, or high-load/mid-speed. This quantile-based regime construction is a pragmatic evaluation device: it creates enough held-out condition cells for paired evaluation while avoiding a manual threshold choice that would be difficult to justify across the evaluated bearings. Because these cutpoints are derived from the frozen analysis set, the regime definition should be read as an evaluation-design choice rather than a deployable train-only operating-condition classifier. Because those regime labels define the held-out test units, they are not provided to the primary model.

\begin{table}[!htbp]
\centering
\caption{Derived load-speed regime coverage in the 10-bearing PHME analysis set.}
\label{tab:regime_counts}
\begin{tabular}{lrrrrr}
\toprule
Regime & Windows & Bearings & Load mean & Speed mean & Life fraction \\
\midrule
low-load / low-speed & 1,526 & 7 & 2683.3 & 1402.2 & 0.00--1.00 \\
low-load / mid-speed & 997 & 3 & 2707.9 & 2342.9 & 0.00--1.00 \\
low-load / high-speed & 2,243 & 4 & 1968.1 & 2702.1 & 0.00--1.00 \\
mid-load / low-speed & 1,766 & 6 & 3179.1 & 1549.1 & 0.00--1.00 \\
mid-load / mid-speed & 1,912 & 8 & 3208.9 & 2388.8 & 0.00--1.00 \\
mid-load / high-speed & 1,087 & 8 & 3364.3 & 2810.9 & 0.01--1.00 \\
high-load / low-speed & 1,474 & 4 & 4310.0 & 1713.2 & 0.03--1.00 \\
high-load / mid-speed & 1,856 & 6 & 4250.4 & 2369.1 & 0.02--1.00 \\
high-load / high-speed & 1,436 & 7 & 4005.8 & 2763.2 & 0.00--1.00 \\
\bottomrule
\end{tabular}

\vspace{0.25em}
{\footnotesize Regimes are derived from load and speed metadata and form the held-out units in leave-operating-regime-out evaluation.}
\end{table}

The regime distribution is not perfectly balanced, which is expected for time-varying operating trajectories. Figure~\ref{fig:regime_distribution} shows the window count per regime, and Figure~\ref{fig:bearing_regime_heatmap} shows which bearings contribute windows to each regime. These figures are used to interpret result variance: a held-out regime with fewer windows or fewer contributing bearings is a more fragile estimate than a broad regime represented across several bearings.

\begin{figure}[!htbp]
\centering
\begin{tikzpicture}[x=1mm,y=1mm]
\fill[blue!59] (25,44) rectangle (59,61);
\draw[white,line width=0.8pt] (25,44) rectangle (59,61);
\node[font=\scriptsize] at (42.0,54.5) {1,474};
\node[font=\tiny] at (42.0,48.5) {4 bearings};
\fill[blue!64] (59,44) rectangle (93,61);
\draw[white,line width=0.8pt] (59,44) rectangle (93,61);
\node[font=\scriptsize] at (76.0,54.5) {1,856};
\node[font=\tiny] at (76.0,48.5) {6 bearings};
\fill[blue!58] (93,44) rectangle (127,61);
\draw[white,line width=0.8pt] (93,44) rectangle (127,61);
\node[font=\scriptsize] at (110.0,54.5) {1,436};
\node[font=\tiny] at (110.0,48.5) {7 bearings};
\fill[blue!63] (25,27) rectangle (59,44);
\draw[white,line width=0.8pt] (25,27) rectangle (59,44);
\node[font=\scriptsize] at (42.0,37.5) {1,766};
\node[font=\tiny] at (42.0,31.5) {6 bearings};
\fill[blue!65] (59,27) rectangle (93,44);
\draw[white,line width=0.8pt] (59,27) rectangle (93,44);
\node[font=\scriptsize] at (76.0,37.5) {1,912};
\node[font=\tiny] at (76.0,31.5) {8 bearings};
\fill[blue!52] (93,27) rectangle (127,44);
\draw[white,line width=0.8pt] (93,27) rectangle (127,44);
\node[font=\scriptsize] at (110.0,37.5) {1,087};
\node[font=\tiny] at (110.0,31.5) {8 bearings};
\fill[blue!59] (25,10) rectangle (59,27);
\draw[white,line width=0.8pt] (25,10) rectangle (59,27);
\node[font=\scriptsize] at (42.0,20.5) {1,526};
\node[font=\tiny] at (42.0,14.5) {7 bearings};
\fill[blue!50] (59,10) rectangle (93,27);
\draw[white,line width=0.8pt] (59,10) rectangle (93,27);
\node[font=\scriptsize] at (76.0,20.5) {997};
\node[font=\tiny] at (76.0,14.5) {3 bearings};
\fill[blue!70] (93,10) rectangle (127,27);
\draw[white,line width=0.8pt] (93,10) rectangle (127,27);
\node[font=\scriptsize] at (110.0,20.5) {2,243};
\node[font=\tiny] at (110.0,14.5) {4 bearings};
\node[font=\scriptsize] at (42.0,6) {low S};
\node[font=\scriptsize] at (76.0,6) {mid S};
\node[font=\scriptsize] at (110.0,6) {high S};
\node[font=\scriptsize,anchor=east] at (22,18.5) {low L};
\node[font=\scriptsize,anchor=east] at (22,35.5) {mid L};
\node[font=\scriptsize,anchor=east] at (22,52.5) {high L};
\node[font=\small] at (76,0) {Speed bin};
\node[font=\small,rotate=90] at (7,35.5) {Load bin};
\end{tikzpicture}

\vspace{0.25em}
{\footnotesize Each cell reports window count and contributing-bearing count. Color intensity follows the square root of the window count so the largest cells do not visually dominate the regime map.}
\caption{Three-by-three derived regime heatmap for the 10-bearing analysis set.}
\label{fig:regime_distribution}
\end{figure}

\begin{figure}[!htbp]
\centering
\resizebox{\textwidth}{!}{%
\small
\setlength{\tabcolsep}{3.4pt}
\renewcommand{\arraystretch}{1.34}
\begin{tabular}{lrrrrrrrrrr}
\toprule
Bearing & low/low & low/mid & low/high & mid/low & mid/mid & mid/high & high/low & high/mid & high/high & Total \\
\midrule
B01 & \cellcolor{blue!40}51 & \cellcolor{gray!8}-- & \cellcolor{gray!8}-- & \cellcolor{gray!8}-- & \cellcolor{blue!41}55 & \cellcolor{blue!45}93 & \cellcolor{gray!8}-- & \cellcolor{blue!42}60 & \cellcolor{blue!47}\textcolor{white}{118} & \textbf{377} \\
B02 & \cellcolor{blue!40}51 & \cellcolor{gray!8}-- & \cellcolor{gray!8}-- & \cellcolor{gray!8}-- & \cellcolor{blue!53}\textcolor{white}{247} & \cellcolor{blue!56}\textcolor{white}{385} & \cellcolor{gray!8}-- & \cellcolor{blue!49}\textcolor{white}{164} & \cellcolor{blue!53}\textcolor{white}{269} & \textbf{1,116} \\
B03 & \cellcolor{gray!8}-- & \cellcolor{gray!8}-- & \cellcolor{gray!8}-- & \cellcolor{gray!8}-- & \cellcolor{blue!49}\textcolor{white}{157} & \cellcolor{blue!51}\textcolor{white}{206} & \cellcolor{gray!8}-- & \cellcolor{blue!46}\textcolor{white}{101} & \cellcolor{blue!49}\textcolor{white}{150} & \textbf{614} \\
B04 & \cellcolor{blue!56}\textcolor{white}{395} & \cellcolor{gray!8}-- & \cellcolor{gray!8}-- & \cellcolor{blue!60}\textcolor{white}{597} & \cellcolor{gray!8}-- & \cellcolor{gray!8}-- & \cellcolor{blue!47}\textcolor{white}{122} & \cellcolor{gray!8}-- & \cellcolor{gray!8}-- & \textbf{1,114} \\
B05 & \cellcolor{blue!47}\textcolor{white}{112} & \cellcolor{blue!42}65 & \cellcolor{blue!38}38 & \cellcolor{blue!46}\textcolor{white}{102} & \cellcolor{blue!48}\textcolor{white}{128} & \cellcolor{blue!43}70 & \cellcolor{blue!33}20 & \cellcolor{blue!36}27 & \cellcolor{blue!28}10 & \textbf{572} \\
B08 & \cellcolor{blue!40}50 & \cellcolor{gray!8}-- & \cellcolor{gray!8}-- & \cellcolor{blue!55}\textcolor{white}{321} & \cellcolor{blue!56}\textcolor{white}{389} & \cellcolor{blue!52}\textcolor{white}{238} & \cellcolor{blue!54}\textcolor{white}{307} & \cellcolor{blue!55}\textcolor{white}{337} & \cellcolor{blue!50}\textcolor{white}{185} & \textbf{1,827} \\
B10 & \cellcolor{blue!56}\textcolor{white}{356} & \cellcolor{blue!56}\textcolor{white}{380} & \cellcolor{blue!38}38 & \cellcolor{blue!43}69 & \cellcolor{blue!44}82 & \cellcolor{blue!30}12 & \cellcolor{blue!64}\textcolor{white}{1025} & \cellcolor{blue!65}\textcolor{white}{1167} & \cellcolor{blue!45}95 & \textbf{3,224} \\
B11 & \cellcolor{blue!58}\textcolor{white}{511} & \cellcolor{blue!59}\textcolor{white}{552} & \cellcolor{blue!39}43 & \cellcolor{blue!55}\textcolor{white}{351} & \cellcolor{blue!57}\textcolor{white}{452} & \cellcolor{blue!39}44 & \cellcolor{gray!8}-- & \cellcolor{gray!8}-- & \cellcolor{gray!8}-- & \textbf{1,953} \\
B12 & \cellcolor{gray!8}-- & \cellcolor{gray!8}-- & \cellcolor{gray!8}-- & \cellcolor{blue!55}\textcolor{white}{326} & \cellcolor{blue!56}\textcolor{white}{402} & \cellcolor{blue!38}39 & \cellcolor{gray!8}-- & \cellcolor{gray!8}-- & \cellcolor{gray!8}-- & \textbf{767} \\
B17 & \cellcolor{gray!8}-- & \cellcolor{gray!8}-- & \cellcolor{blue!70}\textcolor{white}{2124} & \cellcolor{gray!8}-- & \cellcolor{gray!8}-- & \cellcolor{gray!8}-- & \cellcolor{gray!8}-- & \cellcolor{gray!8}-- & \cellcolor{blue!60}\textcolor{white}{609} & \textbf{2,733} \\
\midrule
\textbf{Total} & \textbf{1,526} & \textbf{997} & \textbf{2,243} & \textbf{1,766} & \textbf{1,912} & \textbf{1,087} & \textbf{1,474} & \textbf{1,856} & \textbf{1,436} & \textbf{14,297} \\
\bottomrule
\end{tabular}
}

\vspace{0.25em}
{\footnotesize Cell values are window counts with marginal totals. Darker blue indicates larger log-scaled counts; the transform keeps the largest bearing-regime cell (2124 windows) from dominating the heatmap. Empty cells indicate no available bearing-regime combination in the analysis set. The full grid is retained to document regime imbalance behind the conditional-coverage diagnostics.}
\caption{Bearing-by-regime window-count heatmap in the 10-bearing PHME analysis set.}
\label{fig:bearing_regime_heatmap}
\end{figure}

\subsection{Primary Split: Leave-One-Operating-Regime-Out}

The strict primary endpoint is leave-one-operating-regime-out evaluation with four disjoint roles. For each of the nine derived regimes, all windows in that regime are assigned to the test split. Two non-test regimes are assigned to validation/model selection and empirical residual calibration, respectively, under a shared rotation rule fixed across all model families, and all other regimes are assigned to training. The resulting comparison is therefore regime-disjoint across train, validation, calibration, and test for the held-out operating cell. A matched model-selection/calibration-regime protocol, in which one non-test regime serves both roles, is retained only as a secondary sensitivity analysis.

This split is designed to stress operating-condition generalization, not bearing-identity generalization. Because a single bearing trajectory can pass through multiple regimes, the same bearing identity may appear in training and testing through different load-speed portions of its life. This is acceptable for the central question of this paper, which asks whether models can extrapolate across operating regimes within time-varying bearing data. It would not by itself establish leave-bearing-out generalization or rule out all trajectory-continuity effects. The reported experiments therefore include a separate leave-bearing-out bearing-identity generalization check and a bearing-regime conditional analysis, and treat both as supporting rather than primary. A purged regime split, blocked trajectory split, or joint leave-bearing-and-regime-out split remains a stronger future validation step.

\subsection{Evaluation Blocks}

The experimental suite is organized around evidence blocks that directly support the reliability-oriented framing. The 10-bearing primary comparison evaluates the calibrated predictive-representation model against a random-subspace tabular baseline, a gradient-boosted tabular baseline, a 400-tree random-forest tabular comparator, a temporal-convolutional baseline, and an attention-and-weak-regularization neural baseline. The 10-bearing leave-bearing-out matrix is reported as a bearing-identity generalization check. A context-control diagnostic evaluates no context, load/speed-only context, and load/speed plus regime-code context; this check documents the load/speed-only primary input policy used in the 10-bearing subset study. The diagnostic suite then evaluates ensemble fairness, bounded seed sensitivity, conditional coverage, a pooled regime-conditioned residual diagnostic, bearing-regime decision risk, ablations, prefix-observation behavior, additive-noise stress, raw-channel-loss stress, and high-load/high-speed condition-response behavior.

\begin{table}[!htbp]
\centering
\caption{Evaluation blocks used to support the reliability-oriented evaluation.}
\label{tab:evaluation_protocol}
\resizebox{\textwidth}{!}{%
\small
\begin{tabularx}{\textwidth}{p{0.19\textwidth}p{0.29\textwidth}p{0.14\textwidth}X}
\toprule
Block & Protocol & Unit & Role \\
\midrule
Primary strict endpoint & Leave-one-operating-regime-out with separate train/model-selection/calibration/test regimes & Regime & Main quantitative endpoint. \\
Matched sensitivity & Calibrated predictive-representation model vs. neural and tabular baselines & Regime & Secondary comparison with shared model-selection/calibration regimes. \\
Context-control check & None; load/speed; load/speed + regime & Regime & Design-control analysis that documents load/speed-only primary context. \\
Uncertainty endpoint & Empirical 90\% intervals & Split & Coverage, width, normalized width, and interval score are reported. \\
Diagnostic suite & Prefix observation; additive noise; raw-stream ablation; high-load/high-speed condition response & Regime & 10-bearing diagnostics using the primary load/speed-only policy. \\
Statistics & Bootstrap CI, sign-flip, Wilcoxon & Regime & Primary paired statistics use the strict four-way split. \\
\bottomrule
\end{tabularx}
}

\end{table}

All reported quantitative values are treated as outputs of the documented evaluation pipeline rather than as manually transcribed values. This provenance rule supports reproducibility: every numerical claim should be traceable either to a cited source or to a documented source table or evaluation record.

\section{Method}

\subsection{Problem Definition}

Each supervised example is a vibration window $i$ with raw signal segment $x_i$, engineered features $f_i$, operating context $u_i$, and normalized RUL target $y_i \in [0,1]$. In the primary experiments, $u_i$ contains only normalized measured load and speed. Derived regime labels define held-out splits and reporting strata, but they are not model inputs. The learning goal is a predictor $\hat{y}_i$ that estimates normalized RUL under held-out operating regimes and, after residual calibration, an interval $[\ell_i,h_i]$ with empirical coverage near the nominal 90\% target on the evaluated split.

The setting is retrospective: true failure times are available for labeling and metric computation. The task is therefore offline regime-shift RUL estimation and calibration on a processed subset, not online failure-time discovery.

\subsection{Absolute RUL Conversion}

The model predicts normalized RUL. For conventional absolute-step reporting, each normalized prediction is multiplied by the maximum step count of the corresponding bearing trajectory. This bearing-specific scale is part of retrospective label construction and metric reporting; it is not an inference-time quantity that would be known in a deployed prognostics system before failure. All models and baselines use the same conversion, so the absolute-step MAE comparisons are internally controlled within this offline protocol. To avoid hiding this assumption, the retained prediction files store normalized targets and normalized predictions; the public repository provides code and configuration files to regenerate these outputs where dataset-access constraints permit. The absolute-step results should therefore be read as retrospective PHME-subset metrics, not as a deployment-ready lifetime-normalization procedure.

\subsection{Calibrated Predictive-Representation Predictor}

The main model is a calibrated predictive-representation RUL predictor. The learned representation is calibrated for prediction intervals but is not a high-fidelity physical digital twin. Figure~\ref{fig:framework} summarizes the workflow.

\begin{figure}[!htbp]
\centering
\resizebox{0.96\textwidth}{!}{%
\begin{tikzpicture}[x=1mm,y=1mm,every node/.style={font=\scriptsize},>=latex]
\node[draw,fill=blue!8,minimum width=29mm,minimum height=7mm] (raw) at (15,48) {Raw vibration};
\node[draw,fill=blue!8,minimum width=29mm,minimum height=7mm] (feat) at (15,34) {Engineered feat.};
\node[draw,fill=blue!8,minimum width=29mm,minimum height=7mm] (ctx) at (15,20) {Load/speed};
\node[draw,fill=gray!8,minimum width=29mm,minimum height=7mm] (rawenc) at (52,48) {Signal encoder};
\node[draw,fill=gray!8,minimum width=29mm,minimum height=7mm] (featenc) at (52,34) {Feature encoder};
\node[draw,fill=gray!8,minimum width=29mm,minimum height=7mm] (ctxenc) at (52,20) {Context encoder};
\node[draw,fill=orange!13,minimum width=22mm,minimum height=8mm] (fusion) at (88,34) {Fusion};
\node[draw,fill=teal!12,minimum width=31mm,minimum height=8mm] (latent) at (121,34) {Predictive representation};
\node[draw,fill=green!12,minimum width=25mm,minimum height=8mm] (point) at (153,34) {Point RUL};
\node[draw,dashed,fill=purple!8,minimum width=30mm,minimum height=7mm] (phys) at (121,18) {optional auxiliary losses};
\node[draw,fill=gray!8,minimum width=33mm,minimum height=7mm] (resid) at (70,6) {Calibration residuals};
\node[draw,fill=gray!8,minimum width=29mm,minimum height=7mm] (quant) at (112,6) {Empirical quantile};
\node[draw,fill=red!10,minimum width=30mm,minimum height=8mm] (pi) at (153,6) {Empirical interval};
\draw[->,thick] (raw) -- (rawenc);
\draw[->,thick] (feat) -- (featenc);
\draw[->,thick] (ctx) -- (ctxenc);
\draw[->,thick] (rawenc) -- (fusion);
\draw[->,thick] (featenc) -- (fusion);
\draw[->,thick] (ctxenc) -- (fusion);
\draw[->,thick] (fusion) -- (latent);
\draw[->,thick] (latent) -- (point);
\draw[->,thick,dashed] (latent) -- (phys);
\draw[->,thick] (resid) -- (quant);
\draw[->,thick] (point) -- (pi);
\draw[->,thick] (quant) -- (pi);
\node[font=\tiny,anchor=south] at (91,10.5) {post-hoc residual calibration};
\end{tikzpicture}}

\vspace{0.25em}
{\footnotesize The predictor produces point RUL estimates; prediction intervals are formed after training by empirical residual calibration. Dashed terms are optional auxiliary losses, and the primary model uses load/speed context without derived regime code.}
\caption{Calibrated predictive-representation architecture with explicit post-hoc residual interval calibration.}
\label{fig:framework}
\end{figure}

The model uses three encoders. A one-dimensional convolutional signal encoder maps the raw vibration window to a signal representation,
\begin{equation}
    h^x_i = g_x(x_i).
\end{equation}
An engineered-feature encoder maps the deterministic feature vector to
\begin{equation}
    h^f_i = g_f(f_i),
\end{equation}
and a context encoder maps measured load and speed to
\begin{equation}
    h^u_i = g_u(u_i).
\end{equation}
The fused predictive representation is then
\begin{equation}
    z_i = g_z([h^x_i, h^f_i, h^u_i]).
\end{equation}
In the implemented model, $g_x$ is a three-layer one-dimensional convolutional network with SiLU activations and adaptive average pooling, $g_f$ and $g_u$ are lightweight multilayer perceptrons, and $g_z$ is a two-layer fusion network.

The predictive representation feeds two auxiliary heads: a bounded health-score output and a nonnegative predictive-rate output,
\begin{equation}
    H_i = \sigma(w_H^\top z_i + b_H), \qquad
    r_i = \operatorname{softplus}(w_r^\top z_i + b_r).
\end{equation}
The normalized first-passage RUL surrogate is
\begin{equation}
    \hat{y}_i =
    \frac{(1-H_i)/(r_i+\epsilon)}
         {1 + (1-H_i)/(r_i+\epsilon)} ,
\end{equation}
where $\epsilon=10^{-4}$ in the implementation. This formulation enforces nonnegative predicted RUL and creates an interpretable relationship between the auxiliary health score, the predictive-rate head, and time-to-threshold. It does not by itself guarantee that the learned $H_i$ sequence is monotone along a full bearing trajectory, because each window is encoded independently.

\subsection{Optional Weak Monotonicity and Smoothness Regularization}

Weakly regularized variants are included because the related literature strongly motivates degradation accumulation, first-passage behavior, smoothness, and load-speed effects in RUL modeling \citep{li2024pidd_rul_review,liao2023attnpinn}. The optional regularized objective is
\begin{equation}
    \mathcal{L}
    =
    \frac{1}{n}\sum_i (\hat{y}_i-y_i)^2
    + \lambda \sum_k w_k \mathcal{R}_k ,
\end{equation}
where the regularizers penalize threshold inconsistency, health-score/life misalignment, non-monotone within-bearing health-score sequences, second-difference roughness, and rate behavior that contradicts the weak intuition that higher load or speed should not reduce the learned predictive rate.

These terms are treated as weak inductive biases rather than a complete bearing-fatigue model. This distinction is central to the interpretation. The reported evidence does not show that weak regularization is the mechanism of the accuracy gain, and monotone physical damage is not established by the learned representation.

\subsection{Empirical Residual Interval Calibration}

The point predictor is converted into prediction intervals using empirical residual calibration related to split conformal prediction \citep{lei2018distribution_free,romano2019cqr,angelopoulos2021gentle,javanmardi2023conformal_rul}. Let $\mathcal{C}$ be the calibration set and let $e_i=|y_i-\hat{y}_i|$ be the absolute normalized residual. For nominal miscoverage $\alpha=0.1$, the calibration width is
\begin{equation}
    \hat{q} =
    \operatorname{Quantile}_{\lceil (|\mathcal{C}|+1)(1-\alpha)\rceil / |\mathcal{C}|}
    \{e_i : i \in \mathcal{C}\}.
\end{equation}
The interval for a new prediction is
\begin{equation}
    [\ell_i,h_i] = [\hat{y}_i-\hat{q}, \hat{y}_i+\hat{q}],
\end{equation}
with negative bounds clipped to zero before conversion back to absolute RUL units. Empirical coverage is then measured directly on the held-out split or stress setting.

In the reported primary neural experiments, the calibrated predictive-representation model is a three-member ensemble. The residual quantile is fitted on residuals of the ensemble mean in normalized RUL units, and the reported interval half-width for a test sample is $q_\alpha+s_i$, where $s_i$ is the sample standard deviation across ensemble-member predictions. Thus the reported intervals use both residual calibration and ensemble dispersion. For all multi-member neural rows in the ensemble-fairness table, intervals are computed from the ensemble-mean residual quantile plus member-to-member predictive standard deviation; single-member neural and tabular rows use the residual quantile alone. Both interval endpoints are lower-clipped at zero before conversion to absolute-step units; upper endpoints are not capped at one.

The primary numerical table uses the stricter four-way train/validation/calibration/test split with an explicitly separate calibration regime. The matched model-selection/calibration regime is retained as a sensitivity protocol because it was used in the earlier controlled comparison, but it is not the strictest split-calibration design: the same held-out regime supports early stopping/model selection and residual quantile estimation. Coverage claims are therefore stated as empirical coverage on the evaluated subset rather than finite-sample deployment guarantees.

For conditional-reliability analysis, the manuscript also reports a post-hoc pooled regime-conditioned residual diagnostic. For a test window in regime $g$, this diagnostic estimates the residual radius from calibration residuals belonging to the same load-speed regime $g$ in other strict rotations. The resulting intervals are not the primary reported intervals and are not used to claim a single-split Mondrian conformal guarantee. Their purpose is narrower: to test whether group-specific residual scale explains the observed conditional undercoverage and whether the pattern motivates a pre-specified conditional-calibration experiment.

\subsection{Baselines and Variants}

The evaluation includes five main baselines and comparators. The first tabular baseline is a dependency-free random-subspace ridge ensemble over engineered features and load/speed context. The second tabular baseline is a dependency-free gradient-boosted decision-stump ensemble trained on the same tabular inputs, following the gradient-boosting principle \citep{friedman2001greedy}. The third tabular comparator is a 400-tree random forest over engineered features and load/speed context \citep{breiman2001random_forests}; it is included as a stronger point-error comparator and is not treated as inferior when it has lower MAE. The temporal-convolutional baseline uses dilated one-dimensional convolutions over the raw signal and fuses the resulting representation with engineered features and load/speed context \citep{bai2018tcn}. The scikit-learn gradient-boosted regression and random-forest comparators use the scikit-learn implementation \citep{pedregosa2011scikit}. The attention-and-weak-regularization baseline applies feature attention and predicts RUL together with auxiliary health-score and rate heads under weak monotonicity/smoothness regularization. This name is used deliberately: the implementation is a study-specific attention-plus-regularization comparator inspired by the literature, not a claim to reproduce every detail of a published AttnPINN implementation.

The diagnostic variants separate optional weakly regularized predictive-representation models from component-removal ablations of the primary model. The component ablations remove load/speed context, engineered features, raw vibration, and empirical intervals one factor at a time. The interval-removal ablation is interpreted only as an interval ablation: the point model is unchanged, so MAE should not change and coverage/width are unavailable. These diagnostics are used to interpret the evidence without overstating the role of any single component.

\section{Experimental Setup}

\subsection{Training Protocol}

All neural models are trained from configuration files with fixed seeds and saved evaluation records. The calibrated predictive-representation model uses hidden width 96, a three-member ensemble, AdamW optimization, learning rate $10^{-3}$, weight decay $10^{-4}$, batch size 64, early stopping with patience 18, and up to 120 epochs. Reported neural metrics use the best validation-loss state retained by early stopping, not the last training epoch. Context dropout of 0.5 is applied during training to reduce overdependence on load-speed metadata, and the neural feature/fusion encoders use dropout 0.1. The temporal convolutional network (TCN) and attention-and-weak-regularization baselines use hidden width 96, AdamW optimization, learning rate $10^{-3}$, weight decay $10^{-4}$, batch size 64, early stopping with patience 16, and up to 100 epochs; the attention-and-weak-regularization baseline uses regularization weight 0.05. The random-subspace ridge baseline uses 100 ridge members, ridge penalty $10^{-2}$, and 65\% feature subsampling per member. The dependency-free boosted baseline uses 260 gradient-boosted decision stumps with learning rate 0.035, 32 quantile thresholds per selected feature, and 85\% feature subsampling per boosting step. The scikit-learn gradient-boosted regression comparator uses 400 trees, learning rate 0.04, maximum depth 3, and minimum leaf size 3. The 400-tree random-forest tabular comparator uses minimum leaf size 2 and a single worker process in the evaluation environment. Seeds fix split construction and model initialization; bitwise-identical CUDA replay is not asserted because deterministic CUDA-kernel settings were not enforced in the reported experiments.

All feature and context normalization statistics are estimated on the training split and then reused for held-out splits. The primary strict analysis uses one non-test regime for model selection and a separate non-test regime for empirical residual calibration. Test windows are not used for training, early stopping, model selection, or empirical residual calibration. The primary 10-bearing configurations use load/speed-only context; evaluation records specify that regime labels are used for splitting and reporting but not as model inputs. The matched model-selection/calibration protocol uses a shared non-test regime and is reported only as secondary sensitivity evidence. Both protocols use a shared split seed, so each model family is evaluated on the same held-out test regimes.

\subsection{Primary Comparison}

For each of the nine derived operating regimes, an independent leave-regime-out run is executed for each model family. The main result is the mean over these nine held-out regimes. This design makes each regime a paired evaluation unit, which allows model comparisons to be made by regime rather than by pooling all windows into one large test set.

The candidate in the paired statistical analysis is the calibrated predictive-representation model with load/speed-only context and without optional weak regularization. The primary neural baselines are the TCN and attention-and-weak-regularization models under the same context policy. The tabular comparators include random-subspace ridge, dependency-free boosted decision stumps, and the 400-tree random forest. Weakly regularized predictive-representation variants remain diagnostic variants, but they are not used as the central novelty claim.

\subsection{Metrics}

The primary point-estimation metrics are normalized MAE and normalized RMSE. Absolute-step MAE and RMSE are also reported after applying the common retrospective bearing-specific scale factor, but they are secondary benchmark units rather than deployment-time claims. The retained prediction files store normalized targets and predictions, and the code reports normalized MAE/RMSE for new runs to make the scaling assumption explicit. Uncertainty quality is measured by empirical interval coverage, mean interval width, normalized interval width, and the standard interval score for nominal miscoverage $\alpha=0.1$. Normalized interval width is computed as mean interval width divided by $\max(\operatorname{mean}(|\mathrm{RUL}|),1)$ over the evaluated split; this is why late-stage windows can have large normalized width even when absolute width is lower. The target coverage band for nominal 90\% intervals is 85--95\%, which allows some finite-sample and split-shift variation while still requiring useful calibration. Conditional coverage is reported by load-speed regime and by bearing, and the pooled regime-conditioned calibration diagnostic reports mean regime coverage rather than pooled window coverage. The monotonicity diagnostic is the fraction of negative increments after ordering windows by bearing and step. For predictive-representation models, it is computed from the learned auxiliary health-score sequence. For non-neural baselines, it is computed from the prediction-derived health proxy $1-\hat{y}_{\mathrm{norm}}$. It is reported as a diagnostic only, not as a primary success criterion.

\par\addvspace{0.8\baselineskip}
\noindent\begin{minipage}{\textwidth}
\refstepcounter{table}\label{tab:rul_step_units}
\centering
{\small Table~\thetable: RUL-step unit interpretation for the processed PHME subset.\par}
\vspace{0.25em}
\begin{tabularx}{\textwidth}{p{0.30\textwidth}X}
\toprule
Quantity & Interpretation \\
\midrule
Sampling rate & 25600 Hz \\
Window length & 512 samples \\
Signal duration per window & 0.020 s \\
Window stride & not inferred from the archive; the step index is the ordered PHME window/file index \\
One RUL step & one ordered window/file increment within a bearing trajectory \\
Absolute-step MAE & retrospective benchmark unit after multiplying normalized predictions by each bearing's max step count \\
\bottomrule
\end{tabularx}

\vspace{0.25em}
{\footnotesize Window-level metrics are descriptive because windows are nested within bearing trajectories; inferential comparisons are made over held-out regimes or bearings.}
\end{minipage}
\par\addvspace{0.8\baselineskip}

\par\addvspace{0.8\baselineskip}
\noindent\begin{minipage}{\textwidth}
\refstepcounter{table}\label{tab:normalized_rul_metrics}
\centering
{\small Table~\thetable: Normalized-RUL metrics for the matched-protocol 10-bearing held-out-regime sensitivity evaluation.\par}
\vspace{0.25em}
\begin{tabular}{lrr}
\toprule
Model & Normalized MAE & Normalized RMSE \\
\midrule
400-tree random forest & 0.1409 & 0.1803 \\
Calibrated predictive representation & 0.1410 & 0.1815 \\
Attention-and-weak-regularization baseline & 0.1556 & 0.1949 \\
TCN baseline & 0.1578 & 0.1965 \\
Gradient-boosted tabular & 0.1790 & 0.2149 \\
Random-subspace ridge & 0.1999 & 0.2644 \\
\bottomrule
\end{tabular}

\vspace{0.25em}
{\footnotesize Absolute-step MAE/RMSE in Table~\ref{tab:main_benchmark} use the retrospective bearing-specific lifetime scale. Normalized metrics are computed directly from stored normalized targets and predictions; the strict primary table reports normalized MAE directly.}
\end{minipage}
\par\addvspace{0.8\baselineskip}

For maintenance-risk interpretation, the main decision diagnostic aggregates over bearing-regime groups. A group is late if any saved test window has true normalized RUL at most $\tau=0.20$. For a trigger rule, missed late groups are groups without an on-time trigger at or before the first late window, and false-early groups are groups with no late test window but at least one trigger. The reported stylized loss is
\begin{equation}
    L_\rho = 100(\rho m+f)/n,
\end{equation}
where $m$ is the missed-late count, $f$ is the false-early count, $n$ is the number of bearing-regime groups, and $\rho=c_{\mathrm{miss}}/c_{\mathrm{early}}$ is the miss-to-early-inspection cost ratio. This loss is a retrospective diagnostic for comparing trigger behavior, not a validated maintenance policy.

\subsection{Paired Statistical Tests}

For each baseline comparison, the regime-level MAE improvement is
\begin{equation}
    \Delta_r = \mathrm{MAE}^{\mathrm{baseline}}_r -
               \mathrm{MAE}^{\mathrm{candidate}}_r ,
\end{equation}
where $r$ indexes the held-out regime. The reported relative improvement is $\Delta_r/\mathrm{MAE}^{\mathrm{baseline}}_r$ averaged over regimes. Confidence intervals are obtained by bootstrap resampling the nine paired regime improvements. The sign-flip and Wilcoxon tests are exact for the nine-regime setting. These tests are intentionally paired because all models are evaluated on the same held-out regimes.

\subsection{Stress and Ablation Settings}

Prefix-observation evaluation keeps only test windows up to 20\%, 40\%, 60\%, or 80\% observed life. This creates a stricter early-prognostics view than evaluating only late-life windows. Late-stage evaluation is reported separately for windows with life fraction at least 80\%. Robustness evaluation includes additive signal noise at 0.05 and 0.10 standard deviation in normalized signal units, zeroing the available raw signal channel while retaining precomputed engineered features, and a high-load/high-speed condition-response test that scales the load/speed context. The retained-feature raw-channel-loss row is explicitly a raw-stream ablation, because the precomputed vibration-derived engineered features remain available; stricter feature-degradation sensor-loss probes remain a future experiment outside the reported evidence. Diagnostic variants compare optional weakly regularized predictive-representation models and remove one primary-model component at a time: load/speed context, engineered features, raw vibration, and empirical intervals.

The stress tests are not described as universal robustness guarantees. They are deployment-inspired probes that reveal where the current calibrated predictive-representation framework remains reliable and where it fails.

\section{Results}
\label{sec:results}

We present the evidence in the same order as the reliability argument. First comes the primary strict leave-operating-regime-out endpoint with separate validation and calibration (Table~\ref{tab:calibration_sensitivity}). A matched model-selection/calibration protocol then serves as secondary sensitivity (Table~\ref{tab:main_benchmark}). Conditional coverage and calibration diagnostics follow (Tables~\ref{tab:conditional_regime_coverage}--\ref{tab:regime_conditioned_calibration}). Paired statistics return to the strict endpoint (Table~\ref{tab:paired_statistics}). Architecture, seed, ablation, prefix, and stress checks use the matched diagnostic protocol unless stated otherwise. The final maintenance-risk view returns to the strict split (Table~\ref{tab:bearing_level_decision}).

\subsection{Primary Strict Endpoint and Matched Sensitivity}

Under the strictest four-way split, the first question is whether the calibrated model remains competitive on both error and coverage. Table~\ref{tab:calibration_sensitivity} addresses that question on the 10-bearing PHME analysis set. The calibrated predictive-representation model attains the lowest mean normalized MAE at 0.1477 with empirical 90\% coverage 0.900; its secondary retrospective absolute-step MAE is 285.26. The 400-tree random forest is close on point error (normalized MAE 0.1538, coverage 0.871, absolute-step MAE 294.57), and standard gradient-boosted regression reaches 0.1552, 0.884, and 299.37. The margin is real but modest. The claim supported here is empirical calibration under a bounded regime-disjoint protocol, not broad point-error dominance.

\begin{table}[!htbp]
\centering
\caption{Strict separate-validation/separate-calibration evaluation on the 10-bearing PHME subset.}
\label{tab:calibration_sensitivity}
\resizebox{\textwidth}{!}{%
\begin{tabular}{lrrrrrr}
\toprule
Model & Regimes & Norm. MAE & MAE & Coverage & Width & Interval score \\
\midrule
Calibrated predictive representation & 9 & 0.1477 & 285.26 & 0.900 & 1322.78 & 1709.59 \\
400-tree random forest & 9 & 0.1538 & 294.57 & 0.871 & 1233.14 & 1856.31 \\
Standard gradient-boosted regression & 9 & 0.1552 & 299.37 & 0.884 & 1265.99 & 1695.90 \\
Gradient-boosted tabular & 9 & 0.1816 & 342.28 & 0.908 & 1402.03 & 1560.49 \\
TCN baseline & 9 & 0.1728 & 342.32 & 0.853 & 1259.01 & 2185.33 \\
Random-subspace ridge & 9 & 0.2010 & 384.12 & 0.898 & 1469.21 & 1730.88 \\
Attention-and-weak-regularization baseline & 9 & 0.1919 & 395.84 & 0.841 & 1374.11 & 2898.06 \\
\bottomrule
\end{tabular}
}

\vspace{0.25em}
{\footnotesize For each held-out test regime, model selection and empirical residual calibration use distinct held-out regimes. This stricter four-way train/validation/calibration/test protocol does not extend the scope beyond the ten processed bearings.}
\end{table}

Throughout the primary protocol, models receive measured load and speed only. Derived regime labels define held-out cells for splitting and reporting, but they are withheld as model inputs. The analysis set is the ten processed bearings B01, B02, B03, B04, B05, B08, B10, B11, B12, and B17. This input policy avoids giving models the same derived regime labels used to define held-out evaluation cells; it is a protocol choice rather than a primary performance claim.

The matched protocol in Table~\ref{tab:main_benchmark} asks a softer secondary question: what happens when model selection and residual calibration share a non-test regime? The calibrated model reaches MAE 273.89 and coverage 0.897, inside the 0.85--0.95 band, and remains the lowest-MAE neural row among the TCN and attention-and-weak-regularization baselines (MAE 298.81 and 293.86). The 400-tree random forest again prevents a point-error dominance claim: it has lower mean MAE (266.55) but undercovers at 0.839. Matched sensitivity therefore supports a coverage-oriented neural comparison, not a point-error superiority claim.

\begin{table}[!htbp]
\centering
\caption{Matched-protocol leave-operating-regime-out sensitivity evaluation on the 10-bearing PHME analysis set.}
\label{tab:main_benchmark}
\resizebox{\textwidth}{!}{%
\begin{tabular}{lrrrrrr}
\toprule
Model & Regimes & MAE & RMSE & Coverage & Width & Interval score \\
\midrule
400-tree random forest & 9 & 266.55 & 353.22 & 0.839 & 1209.67 & 1939.65 \\
Calibrated predictive representation & 9 & 273.89 & 362.52 & 0.897 & 1308.82 & 1859.90 \\
Attention-and-weak-regularization baseline & 9 & 293.86 & 379.41 & 0.859 & 1230.55 & 1771.59 \\
TCN baseline & 9 & 298.81 & 379.91 & 0.859 & 1292.12 & 1858.85 \\
Gradient-boosted tabular & 9 & 339.12 & 428.29 & 0.834 & 1325.29 & 1867.10 \\
Random-subspace ridge & 9 & 380.26 & 592.45 & 0.809 & 1269.74 & 1820.34 \\
\bottomrule
\end{tabular}
}

\vspace{0.25em}
{\footnotesize All rows use the load/speed-only context policy and a shared model-selection/calibration regime for each held-out test regime. The 400-tree random forest is a stronger tabular comparator; derived regime labels are used for split construction and reporting, not as model inputs.}
\end{table}

The uncertainty result should be read as empirical coverage on the evaluated endpoint, not as uniformly superior interval quality. Under the matched protocol, the calibrated predictive-representation interval score is 1859.90, close to the TCN and boosted-tabular scores but higher than the attention-and-weak-regularization and random-subspace tabular scores; the random-forest interval score is also worse at 1939.65. In normalized RUL units, random forest and the calibrated predictive-representation model are nearly tied in MAE, 0.1409 versus 0.1410, but random forest remains below the empirical coverage target. This motivates the later stress and limitation analysis: the calibrated model reaches target-band coverage on the primary endpoint, but sharper calibration remains a future improvement.

\subsection{Conditional Reliability and Calibration Diagnostics}

Average coverage can hide local failure. Table~\ref{tab:conditional_regime_coverage} and Figure~\ref{fig:conditional_coverage_heatmap} break the strict primary intervals down by load--speed regime. Two fragilities stand out: low-load/high-speed is the hardest reported-interval cell, and several cells are dominated by a single bearing trajectory. Approximate binomial 95\% confidence intervals are included so that point estimates are not over-read. Table~\ref{tab:conditional_bearing_coverage} pools coverage by bearing across held-out-regime tests; Table~\ref{tab:dominant_bearing_jackknife} removes the dominant bearing in concentrated cells as a row-level diagnostic; and Table~\ref{tab:calibration_curve} checks nominal-versus-empirical coverage across several nominal levels rather than only 90\%.

\begin{table}[!htbp]
\centering
\caption{Per-regime conditional coverage for the separate-calibrated predictive-representation model.}
\label{tab:conditional_regime_coverage}
\resizebox{\textwidth}{!}{%
\begin{tabular}{lrrrrrrr}
\toprule
Regime & Windows & Bearings & Dominant share & Norm. MAE & Coverage & 95\% CI & Interval score \\
\midrule
low-load / low-speed & 1,526 & 7 & 0.335 & 0.1122 & 0.971 & 0.961--0.978 & 1248.12 \\
low-load / mid-speed & 997 & 3 & 0.554 & 0.1108 & 0.973 & 0.961--0.981 & 1579.20 \\
low-load / high-speed & 2,243 & 4 & 0.947 & 0.2595 & 0.666 & 0.646--0.685 & 4419.81 \\
mid-load / low-speed & 1,766 & 6 & 0.338 & 0.1424 & 0.904 & 0.889--0.917 & 962.97 \\
mid-load / mid-speed & 1,912 & 8 & 0.236 & 0.1674 & 0.815 & 0.797--0.832 & 1057.90 \\
mid-load / high-speed & 1,087 & 8 & 0.354 & 0.1476 & 0.949 & 0.935--0.961 & 887.41 \\
high-load / low-speed & 1,474 & 4 & 0.695 & 0.1180 & 1.000 & 0.997--1.000 & 2683.76 \\
high-load / mid-speed & 1,856 & 6 & 0.629 & 0.1241 & 0.912 & 0.898--0.924 & 1345.02 \\
high-load / high-speed & 1,436 & 7 & 0.424 & 0.1470 & 0.914 & 0.899--0.928 & 1202.07 \\
\bottomrule
\end{tabular}
}

\vspace{0.25em}
{\footnotesize Dominant share is the fraction of windows in the held-out regime contributed by the largest single bearing. The confidence interval is an approximate Wilson binomial interval over windows and is descriptive because windows are nested within bearing trajectories.}
\end{table}

Thus the nominal-90\% result should be interpreted as average empirical calibration across the nine held-out regimes, not as uniform conditional calibration.

\begin{figure}[!htbp]
\centering
\begin{center}
\begin{tikzpicture}[x=0.27\textwidth,y=1.05cm]
\node[anchor=east,font=\small] at (-0.10,0.35) {High load};
\draw[fill=blue!16,draw=gray!70] (0,0) rectangle (1,0.7);
\node[font=\small] at (0.5,0.46) {O 1.000};
\node[font=\tiny] at (0.5,0.19) {N=1,474, max=0.70};
\draw[fill=green!18,draw=gray!70] (1,0) rectangle (2,0.7);
\node[font=\small] at (1.5,0.46) {T 0.912};
\node[font=\tiny] at (1.5,0.19) {N=1,856, max=0.63};
\draw[fill=green!18,draw=gray!70] (2,0) rectangle (3,0.7);
\node[font=\small] at (2.5,0.46) {T 0.914};
\node[font=\tiny] at (2.5,0.19) {N=1,436, max=0.42};
\node[anchor=east,font=\small] at (-0.10,1.35) {Mid load};
\draw[fill=green!18,draw=gray!70] (0,1) rectangle (1,1.7);
\node[font=\small] at (0.5,1.46) {T 0.904};
\node[font=\tiny] at (0.5,1.19) {N=1,766, max=0.34};
\draw[fill=red!18,draw=gray!70] (1,1) rectangle (2,1.7);
\node[font=\small] at (1.5,1.46) {U 0.815};
\node[font=\tiny] at (1.5,1.19) {N=1,912, max=0.24};
\draw[fill=green!18,draw=gray!70] (2,1) rectangle (3,1.7);
\node[font=\small] at (2.5,1.46) {T 0.949};
\node[font=\tiny] at (2.5,1.19) {N=1,087, max=0.35};
\node[anchor=east,font=\small] at (-0.10,2.35) {Low load};
\draw[fill=blue!16,draw=gray!70] (0,2) rectangle (1,2.7);
\node[font=\small] at (0.5,2.46) {O 0.971};
\node[font=\tiny] at (0.5,2.19) {N=1,526, max=0.33};
\draw[fill=blue!16,draw=gray!70] (1,2) rectangle (2,2.7);
\node[font=\small] at (1.5,2.46) {O 0.973};
\node[font=\tiny] at (1.5,2.19) {N=997, max=0.55};
\draw[fill=red!18,draw=gray!70] (2,2) rectangle (3,2.7);
\node[font=\small] at (2.5,2.46) {U 0.666};
\node[font=\tiny] at (2.5,2.19) {N=2,243, max=0.95};
\node[anchor=south,font=\small] at (0.5,2.82) {Low speed};
\node[anchor=south,font=\small] at (1.5,2.82) {Mid speed};
\node[anchor=south,font=\small] at (2.5,2.82) {High speed};
\end{tikzpicture}
\end{center}

\vspace{0.25em}
{\footnotesize Cell text gives status plus empirical coverage: U denotes undercoverage below 0.85, T denotes the 0.85--0.95 target band, and O denotes overcoverage above 0.95. The smaller line gives window count and dominant-bearing share, so the display does not rely on color alone.}
\caption{Per-regime conditional coverage heatmap for the separate-calibrated predictive-representation model.}
\label{fig:conditional_coverage_heatmap}
\end{figure}

\begin{table}[!htbp]
\centering
\caption{Dominant-bearing exclusion diagnostic for the separate-calibrated predictive-representation model.}
\label{tab:dominant_bearing_jackknife}
\resizebox{\textwidth}{!}{%
\begin{tabular}{lrrrrr}
\toprule
Regime & Dominant bearing & Share & Coverage & Excl. dominant & N excl. \\
\midrule
low-load / low-speed & B11 & 0.335 & 0.971 & 0.967 & 1,015 \\
low-load / mid-speed & B11 & 0.554 & 0.973 & 0.969 & 445 \\
low-load / high-speed & B17 & 0.947 & 0.666 & 0.840 & 119 \\
mid-load / low-speed & B04 & 0.338 & 0.904 & 0.862 & 1,169 \\
mid-load / mid-speed & B11 & 0.236 & 0.815 & 0.769 & 1,460 \\
mid-load / high-speed & B02 & 0.354 & 0.949 & 0.933 & 702 \\
high-load / low-speed & B10 & 0.695 & 1.000 & 1.000 & 449 \\
high-load / mid-speed & B10 & 0.629 & 0.912 & 0.859 & 689 \\
high-load / high-speed & B17 & 0.424 & 0.914 & 0.851 & 827 \\
\bottomrule
\end{tabular}
}

\vspace{0.25em}
{\footnotesize Coverage excluding the dominant bearing is a row-level diagnostic computed from saved test predictions; no model is retrained. Empty or small excluded sets should be read as evidence of regime-bearing confounding, not as independent regime-level validation.}
\end{table}

\begin{table}[!htbp]
\centering
\caption{Per-bearing conditional coverage across the held-out-regime tests for the separate-calibrated predictive-representation model.}
\label{tab:conditional_bearing_coverage}
\resizebox{\textwidth}{!}{%
\begin{tabular}{lrrrrr}
\toprule
Bearing & Regimes seen & Windows & Norm. MAE & Coverage & Interval score \\
\midrule
B01 & 5 & 377 & 0.1223 & 0.950 & 275.60 \\
B02 & 5 & 1,116 & 0.1358 & 0.917 & 774.86 \\
B03 & 4 & 614 & 0.1920 & 0.785 & 581.66 \\
B04 & 3 & 1,114 & 0.0964 & 0.985 & 728.39 \\
B05 & 9 & 572 & 0.2298 & 0.750 & 727.73 \\
B08 & 7 & 1,827 & 0.1343 & 0.922 & 1314.58 \\
B10 & 9 & 3,224 & 0.1158 & 0.966 & 2327.85 \\
B11 & 6 & 1,953 & 0.1162 & 0.976 & 1245.57 \\
B12 & 3 & 767 & 0.2365 & 0.671 & 1006.52 \\
B17 & 2 & 2,733 & 0.2303 & 0.733 & 3876.42 \\
\bottomrule
\end{tabular}
}

\vspace{0.25em}
{\footnotesize Rows pool each bearing's windows across the nine held-out-regime test predictions; they are conditional diagnostics, not independent bearing-level hypothesis tests.}
\end{table}

\FloatBarrier

\begin{table}[!htbp]
\centering
\caption{Residual-only calibration diagnostic under separate calibration, not primary reported interval coverage.}
\label{tab:calibration_curve}
\begin{tabular}{lrrr}
\toprule
Nominal coverage & Predictive representation & Random forest & TCN \\
\midrule
0.50 & 0.541 & 0.511 & 0.458 \\
0.70 & 0.722 & 0.708 & 0.642 \\
0.80 & 0.796 & 0.794 & 0.744 \\
0.85 & 0.829 & 0.831 & 0.796 \\
0.90 & 0.868 & 0.869 & 0.852 \\
0.95 & 0.918 & 0.922 & 0.904 \\
\bottomrule
\end{tabular}

\vspace{0.25em}
{\footnotesize Each entry averages empirical test coverage over the nine held-out regimes after recalculating split-specific residual-only radii with plain empirical residual quantiles from the separate calibration regime. These values are not the primary reported interval coverage in the main benchmark table, and this diagnostic is not the ensemble-dispersion interval used for the primary reported rows.}
\end{table}

\FloatBarrier

\begin{table}[!htbp]
\centering
\caption{Conditional-reliability diagnostic for alternative empirical calibration rules.}
\label{tab:regime_conditioned_calibration}
\small
\begin{tabularx}{\textwidth}{p{0.30\textwidth}XXXXX}
\toprule
Calibration diagnostic & Mean cov. & Min cov. & LL/HS cov. & Mean norm. width & Fallback rows \\
\midrule
Primary reported interval & 0.900 & 0.666 & 0.666 & 0.653 & -- \\
Global residual-only empirical & 0.869 & 0.609 & 0.609 & 0.597 & -- \\
Pooled regime-conditioned empirical & 0.918 & 0.887 & 0.941 & 0.612 & 0 \\
\bottomrule
\end{tabularx}

\vspace{0.25em}
{\footnotesize Target coverage is nominal 0.90, and coverage/width are averaged over held-out regimes rather than pooled over windows. LL/HS denotes the low-load/high-speed regime. Mean interval score is omitted here to keep the conditional-coverage diagnostic readable. The primary reported interval is the strict-split interval used in the main result tables. Residual-only rows recompute symmetric empirical radii in normalized RUL units. The pooled regime-conditioned row uses calibration residuals from other strict rotations with the same load-speed regime; it is a post-hoc conditional-reliability diagnostic, not a replacement for the primary reported interval or a single-split Mondrian conformal guarantee.}
\end{table}

\FloatBarrier

Table~\ref{tab:regime_conditioned_calibration} adds a reliability-oriented conditional calibration diagnostic. The primary reported interval reaches 0.900 mean regime coverage, but its minimum regime coverage remains 0.666 in low-load/high-speed. The pooled regime-conditioned diagnostic reaches 0.918 mean regime coverage and 0.941 in low-load/high-speed with mean normalized width 0.612. Because it pools calibration residuals across rotations, it is evidence for future group-conditioned calibration, not a primary reported interval or single-split guarantee.

\FloatBarrier

\subsection{Strict-Endpoint Paired Statistical Evidence}

Table~\ref{tab:paired_statistics} gives the paired regime-level statistical comparison for the strict four-way endpoint. Against the TCN baseline, the calibrated predictive-representation model improves mean MAE by 10.38\%, with a 95\% bootstrap relative-improvement confidence interval of 0.83--19.31\%, and has lower MAE in eight of nine regimes; the exact sign-flip test is borderline at 0.0586 and the Wilcoxon test is 0.0391. Against the attention-and-weak-regularization baseline, the relative improvement is 13.41\%, with a 95\% bootstrap interval of 3.67--25.15\%, lower MAE in eight of nine regimes, and exact sign-flip and Wilcoxon tests of 0.0273. Against the standard scikit-learn gradient-boosted regression comparator, the improvement is smaller and uncertain, 7.52\% with a confidence interval of -0.77--14.75\%. Against the 400-tree random forest, the improvement is only 3.89\%, the confidence interval crosses zero, and the model has lower MAE in five of nine regimes. The strict endpoint therefore supports a calibrated neural reference with positive evidence against the neural baselines, but not decisive point-error superiority over the strongest tabular comparators.

\begin{table}[!htbp]
\centering
\caption{Strict-endpoint paired regime-level statistical comparison for the calibrated predictive-representation model.}
\label{tab:paired_statistics}
\resizebox{\textwidth}{!}{%
\begin{tabular}{lrrrrrrr}
\toprule
Baseline & Candidate MAE & Baseline MAE & Rel. gain & 95\% CI & Wins & Sign $p$ & Wilcoxon $p$ \\
\midrule
TCN baseline & 285.26 & 342.32 & 10.38\% & [0.83\%, 19.31\%] & 8/9 & 0.0586 & 0.0391 \\
Attention-and-weak-regularization baseline & 285.26 & 395.84 & 13.41\% & [3.67\%, 25.15\%] & 8/9 & 0.0273 & 0.0273 \\
400-tree random forest & 285.26 & 294.57 & 3.89\% & [-5.66\%, 13.31\%] & 5/9 & 0.4688 & 0.6523 \\
Standard gradient-boosted regression & 285.26 & 299.37 & 7.52\% & [-0.77\%, 14.75\%] & 6/9 & 0.3672 & 0.2500 \\
Gradient-boosted tabular & 285.26 & 342.28 & 19.38\% & [4.71\%, 31.65\%] & 8/9 & 0.1328 & 0.0977 \\
Random-subspace ridge & 285.26 & 384.12 & 27.50\% & [10.77\%, 40.96\%] & 8/9 & 0.0469 & 0.0977 \\
\bottomrule
\end{tabular}
}

\vspace{0.25em}
{\footnotesize Bootstrap relative-improvement confidence intervals, exact sign-flip tests, and Wilcoxon tests are computed over the nine held-out regimes under the strict train/validation/calibration/test split. The 400-tree random-forest row is included because this stronger tabular comparator is close and is not dominated on point error.}
\end{table}

\FloatBarrier

The matched-validation sensitivity analysis remains secondary: it gives a useful same-regime model-selection/calibration comparison, but it is not the main statistical endpoint. The conservative interpretation is unchanged: the calibrated predictive-representation model is a calibrated neural reference, while the stronger random forest remains close enough that these results do not establish general point-error dominance.

Figure~\ref{fig:per_regime_mae} shows the strict-endpoint mean MAE--coverage tradeoff and per-regime relative MAE gains against the TCN, attention-and-weak-regularization, and 400-tree random-forest comparators. The calibrated predictive-representation model loses to TCN in the high-load/high-speed regime, to the attention-and-weak-regularization baseline in mid-load/mid-speed, and to the random forest in four of nine regimes; therefore, the evidence does not establish uniform dominance. The controlled evidence supports competitive calibrated neural performance. Paired inference remains limited by the nine-regime sample size.

\begin{figure}[!htbp]
\centering
\resizebox{0.96\textwidth}{!}{%
\begin{tikzpicture}
\begin{groupplot}[
group style={group size=1 by 2,vertical sep=30mm},
width=\textwidth,
height=50mm,
tick label style={font=\scriptsize},
label style={font=\small},
title style={font=\small},
ymajorgrids=true,
grid style={gray!20},
axis line style={gray!60},
]
\nextgroupplot[
title={(a) Strict-endpoint mean MAE--coverage tradeoff},
xmin=250,xmax=420,
ymin=0.75,ymax=0.95,
xlabel={Mean absolute-step MAE},
ylabel={Empirical coverage},
clip=false,
]
\addplot[draw=none,fill=green!18] coordinates {(250,0.85) (420,0.85) (420,0.95) (250,0.95)} -- cycle;
\addplot+[dashed,draw=gray!70,mark=none] coordinates {(250,0.90) (420,0.90)};
\addplot+[only marks,mark=*,mark size=3.0pt,draw=blue!70!black,fill=blue!70!black] coordinates {(285.26,0.900)};
\addplot+[only marks,mark=triangle*,mark size=3.0pt,draw=red!70!black,fill=red!70!black] coordinates {(294.57,0.871)};
\addplot+[only marks,mark=square*,mark size=3.0pt,draw=orange!80!black,fill=orange!80!black] coordinates {(395.84,0.841)};
\addplot+[only marks,mark=diamond*,mark size=3.0pt,draw=teal!70!black,fill=teal!70!black] coordinates {(342.32,0.853)};
\addplot+[only marks,mark=o,mark size=3.0pt,draw=purple!70!black,fill=purple!70!black] coordinates {(342.28,0.908)};
\addplot+[only marks,mark=x,mark size=3.0pt,draw=gray!75!black,fill=gray!75!black] coordinates {(384.12,0.898)};
\node[font=\scriptsize,anchor=south west,xshift=2pt,yshift=2pt,fill=white,inner sep=0.9pt,text=blue!70!black] at (axis cs:285.26,0.900) {1};
\node[font=\scriptsize,anchor=north west,xshift=2pt,yshift=-2pt,fill=white,inner sep=0.9pt,text=red!70!black] at (axis cs:294.57,0.871) {2};
\node[font=\scriptsize,anchor=south east,xshift=-2pt,yshift=2pt,fill=white,inner sep=0.9pt,text=orange!80!black] at (axis cs:395.84,0.841) {3};
\node[font=\scriptsize,anchor=north east,xshift=-2pt,yshift=-2pt,fill=white,inner sep=0.9pt,text=teal!70!black] at (axis cs:342.32,0.853) {4};
\node[font=\scriptsize,anchor=south west,xshift=2pt,yshift=2pt,fill=white,inner sep=0.9pt,text=purple!70!black] at (axis cs:342.28,0.908) {5};
\node[font=\scriptsize,anchor=south,yshift=-3pt,fill=white,inner sep=0.9pt,text=gray!75!black] at (axis cs:384.12,0.898) {6};
\node[font=\tiny,anchor=south west,text=gray!65!black,fill=white,inner sep=0.8pt] at (axis cs:252,0.755) {$\leftarrow$ lower MAE better};
\node[font=\scriptsize,anchor=north,align=center,fill=white,draw=gray!35,inner sep=1.6pt] at (rel axis cs:0.5,-0.28) {%
1=Predictive representation\quad
2=400-tree forest\quad
3=Attn.+weak-reg.\\[0.3em]
4=TCN\quad
5=Boosted tabular\quad
6=Subspace ridge};

\nextgroupplot[
title={(b) Strict-endpoint relative MAE gain by held-out regime},
xmin=-25,xmax=60,
xlabel={Relative MAE gain (\%)},
ylabel={Held-out regime},
ymin=0.5,ymax=9.5,
ytick={1,2,3,4,5,6,7,8,9},
yticklabels={low L / low S,low L / mid S,low L / high S,mid L / low S,mid L / mid S,mid L / high S,high L / low S,high L / mid S,high L / high S},
yticklabel style={font=\scriptsize},
legend style={at={(0.5,-0.30)},anchor=north,legend columns=3,font=\scriptsize,draw=none},
]
\addplot+[mark=none,dashed,draw=gray!70] coordinates {(0,0.5) (0,9.5)};
\addplot+[mark=none,draw=teal!70!black,line width=0.45pt] coordinates {(0,0.78) (17.62,0.78)};
\addplot+[mark=none,draw=teal!70!black,line width=0.45pt] coordinates {(0,1.78) (7.16,1.78)};
\addplot+[mark=none,draw=teal!70!black,line width=0.45pt] coordinates {(0,2.78) (27.27,2.78)};
\addplot+[mark=none,draw=teal!70!black,line width=0.45pt] coordinates {(0,3.78) (5.24,3.78)};
\addplot+[mark=none,draw=teal!70!black,line width=0.45pt] coordinates {(0,4.78) (6.19,4.78)};
\addplot+[mark=none,draw=teal!70!black,line width=0.45pt] coordinates {(0,5.78) (0.10,5.78)};
\addplot+[mark=none,draw=teal!70!black,line width=0.45pt] coordinates {(0,6.78) (31.39,6.78)};
\addplot+[mark=none,draw=teal!70!black,line width=0.45pt] coordinates {(0,7.78) (17.13,7.78)};
\addplot+[mark=none,draw=teal!70!black,line width=0.45pt] coordinates {(0,8.78) (-18.69,8.78)};
\addplot+[mark=none,draw=orange!80!black,line width=0.45pt] coordinates {(0,1.00) (4.71,1.00)};
\addplot+[mark=none,draw=orange!80!black,line width=0.45pt] coordinates {(0,2.00) (13.50,2.00)};
\addplot+[mark=none,draw=orange!80!black,line width=0.45pt] coordinates {(0,3.00) (52.75,3.00)};
\addplot+[mark=none,draw=orange!80!black,line width=0.45pt] coordinates {(0,4.00) (10.55,4.00)};
\addplot+[mark=none,draw=orange!80!black,line width=0.45pt] coordinates {(0,5.00) (-9.25,5.00)};
\addplot+[mark=none,draw=orange!80!black,line width=0.45pt] coordinates {(0,6.00) (5.16,6.00)};
\addplot+[mark=none,draw=orange!80!black,line width=0.45pt] coordinates {(0,7.00) (17.95,7.00)};
\addplot+[mark=none,draw=orange!80!black,line width=0.45pt] coordinates {(0,8.00) (0.12,8.00)};
\addplot+[mark=none,draw=orange!80!black,line width=0.45pt] coordinates {(0,9.00) (25.20,9.00)};
\addplot+[mark=none,draw=red!80!black,line width=0.45pt] coordinates {(0,1.22) (11.41,1.22)};
\addplot+[mark=none,draw=red!80!black,line width=0.45pt] coordinates {(0,2.22) (-0.55,2.22)};
\addplot+[mark=none,draw=red!80!black,line width=0.45pt] coordinates {(0,3.22) (-3.81,3.22)};
\addplot+[mark=none,draw=red!80!black,line width=0.45pt] coordinates {(0,4.22) (12.03,4.22)};
\addplot+[mark=none,draw=red!80!black,line width=0.45pt] coordinates {(0,5.22) (-15.54,5.22)};
\addplot+[mark=none,draw=red!80!black,line width=0.45pt] coordinates {(0,6.22) (27.29,6.22)};
\addplot+[mark=none,draw=red!80!black,line width=0.45pt] coordinates {(0,7.22) (6.50,7.22)};
\addplot+[mark=none,draw=red!80!black,line width=0.45pt] coordinates {(0,8.22) (-20.23,8.22)};
\addplot+[mark=none,draw=red!80!black,line width=0.45pt] coordinates {(0,9.22) (17.94,9.22)};
\addplot+[only marks,mark=*,mark size=2.7pt,draw=teal!70!black,fill=teal!70!black] coordinates {(17.61893070806665,0.78) (7.156702829507738,1.78) (27.26525538534513,2.78) (5.23621327954865,3.78) (6.193896652101805,4.78) (0.10418790243462313,5.78) (31.39038674591066,6.78) (17.134184309916545,7.78) (-18.68934767946133,8.78)};\addlegendentry{TCN}
\addplot+[only marks,mark=square*,mark size=2.7pt,draw=orange!80!black,fill=orange!80!black] coordinates {(4.710780377849512,1.00) (13.49650846950816,2.00) (52.75272764073144,3.00) (10.546340321808186,4.00) (-9.254727875911437,5.00) (5.156817102623188,6.00) (17.945290155746964,7.00) (0.11855664769921762,8.00) (25.19827326830772,9.00)};\addlegendentry{Attn.+weak-reg.}
\addplot+[only marks,mark=triangle*,mark size=3.4pt,draw=red!80!black,fill=red!80!black] coordinates {(11.41050082193588,1.22) (-0.5474966690022572,2.22) (-3.809388745260213,3.22) (12.027295672782975,4.22) (-15.540377253787987,5.22) (27.289840257846766,6.22) (6.502311941316513,7.22) (-20.234259888355943,8.22) (17.938564319412,9.22)};\addlegendentry{400-tree forest}
\end{groupplot}
\end{tikzpicture}
}

\vspace{0.25em}
{\footnotesize Both panels use the strict four-way train/validation/calibration/test endpoint. Panel (a) plots absolute-step mean MAE and empirical coverage from Table~\ref{tab:calibration_sensitivity}; the shaded band marks the predefined 0.85--0.95 coverage target. Numbered markers identify comparator labels without direct text overlap. Panel (b) reports relative MAE gain by held-out regime versus TCN, attention-and-weak-regularization, and 400-tree random-forest comparators; positive values indicate regimes where the predictive representation has lower MAE. The random forest is included because it is the strongest tabular point-error competitor and is not dominated on MAE.}
\caption{Strict-endpoint MAE--coverage tradeoff and per-regime relative MAE gains against the three main comparators.}
\label{fig:per_regime_mae}
\end{figure}

\FloatBarrier

\subsection{Matched-Protocol Architecture and Seed Diagnostics}

The following architecture, seed, ablation, prefix-observation, and stress diagnostics use the matched model-selection/calibration protocol in Table~\ref{tab:main_benchmark} and are diagnostic rather than primary-endpoint results.

Table~\ref{tab:ensemble_fairness} separates architecture, ensemble, and comparator effects under this matched diagnostic protocol. The three-member calibrated predictive-representation ensemble improves over its single-model version, but the gain is modest. Ensemble-matched TCN and attention-and-weak-regularization baselines narrow the comparison but do not overturn the calibrated-neural interpretation. The strongest point-error comparator in this table is the 400-tree random forest, which reaches MAE 266.55 but empirical coverage 0.839; the paper therefore treats it as a serious tabular competitor rather than as a dominated baseline.

\begin{table}[!htbp]
\centering
\caption{Matched-protocol architecture, ensemble, and input-stream fairness diagnostics.}
\label{tab:ensemble_fairness}
\resizebox{\textwidth}{!}{%
\begin{tabular}{llrrrr}
\toprule
Model & Members/input & MAE & Coverage & Interval score & Monotonicity viol. \\
\midrule
Calibrated predictive representation & 3; raw + features + load/speed & 273.89 & 0.897 & 1859.90 & 0.494 \\
Predictive-representation single model & 1; raw + features + load/speed & 280.21 & 0.867 & 1917.47 & 0.493 \\
TCN baseline & 1; raw + features + load/speed & 298.81 & 0.859 & 1858.85 & 0.500 \\
TCN 3-member ensemble & 3; raw + features + load/speed & 295.61 & 0.889 & 1844.22 & 0.497 \\
Attention-and-weak-regularization baseline & 1; features + load/speed & 293.86 & 0.859 & 1771.59 & 0.502 \\
Attention-and-weak-regularization ensemble & 3; features + load/speed & 291.36 & 0.888 & 1760.49 & 0.492 \\
400-tree random forest & 400 trees; features + load/speed & 266.55 & 0.839 & 1939.65 & 0.494 \\
Feature-only MLP & 1; features + load/speed & 298.24 & 0.856 & 1988.85 & 0.499 \\
Raw-only TCN & 1; raw + load/speed & 303.99 & 0.841 & 2198.22 & 0.497 \\
\bottomrule
\end{tabular}
}

\vspace{0.25em}
{\footnotesize These diagnostics use the matched model-selection/calibration protocol rather than the strict primary endpoint. The 400-tree random forest has the lowest mean MAE but empirical coverage below the target band. The monotonicity diagnostic is computed from learned auxiliary health-score predictions when available and otherwise from the prediction-derived health proxy $1-\hat{y}_{\mathrm{norm}}$ after ordering windows by bearing and step.}
\end{table}

Table~\ref{tab:seed_sensitivity} reports a bounded seed-sensitivity diagnostic over three representative held-out regimes and three model seeds. This is not a full five-seed, nine-regime robustness campaign. It is included to expose seed-scale variation within the limited replication design: on this deliberately difficult three-regime diagnostic, the attention-and-weak-regularization comparator has lower mean MAE, while the calibrated predictive-representation model has lower seed-to-seed MAE variation; therefore, the seed diagnostic supports stability but not uniform neural dominance. The difficult low-load/high-speed cell depresses coverage for all three neural families.

\begin{table}[!htbp]
\centering
\caption{Bounded seed-sensitivity diagnostic on three representative held-out regimes.}
\label{tab:seed_sensitivity}
\resizebox{\textwidth}{!}{%
\begin{tabular}{lrrrr}
\toprule
Model & Regimes $\times$ seeds & MAE mean & MAE SD & Coverage mean \\
\midrule
Predictive representation & 9 & 379.05 & 7.26 & 0.791 \\
TCN & 9 & 386.23 & 23.41 & 0.781 \\
Attention-and-weak-regularization & 9 & 346.54 & 20.26 & 0.812 \\
\bottomrule
\end{tabular}
}

\vspace{0.25em}
{\footnotesize The diagnostic uses three seeds on high-load/high-speed, low-load/high-speed, and mid-load/mid-speed regimes. It is not a full 5-seed, nine-regime campaign, but it shows that the difficult low-load/high-speed cell dominates seed-level variability.}
\end{table}

\subsection{Leave-Bearing-Out Generalization Check}

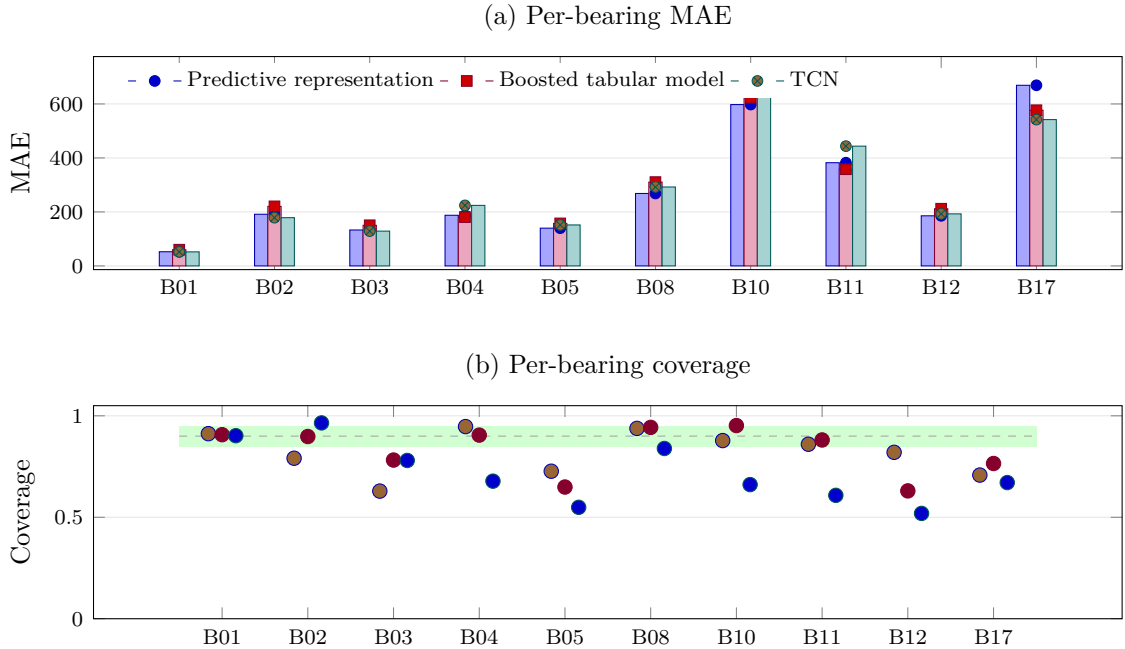
\begin{figure}[!htbp]
\centering
\begin{tikzpicture}
\begin{groupplot}[group style={group size=1 by 2,vertical sep=18mm},width=0.92\textwidth,height=44mm,tick label style={font=\scriptsize},label style={font=\small},title style={font=\small},legend style={at={(0.03,0.97)},anchor=north west,legend columns=3,font=\scriptsize,draw=none,fill=white,inner sep=1pt},ymajorgrids=true,grid style={gray!20}]
\nextgroupplot[title={(a) Per-bearing MAE},ylabel={MAE},xtick={1,2,3,4,5,6,7,8,9,10},xticklabels={B01,B02,B03,B04,B05,B08,B10,B11,B12,B17},xticklabel style={font=\scriptsize}]
\addplot+[ybar,bar shift=-5pt,bar width=5.0pt,draw=blue!70!black,fill=blue!35] coordinates {(1,52.51) (2,191.36) (3,133.07) (4,187.60) (5,139.84) (6,268.35) (7,597.72) (8,382.31) (9,185.61) (10,669.10)};\addlegendentry{Predictive representation}
\addplot+[ybar,bar shift=0pt,bar width=5.0pt,draw=purple!70!black,fill=purple!35] coordinates {(1,60.06) (2,220.57) (3,150.52) (4,180.96) (5,157.46) (6,310.45) (7,623.11) (8,358.26) (9,211.75) (10,576.68)};\addlegendentry{Boosted tabular model}
\addplot+[ybar,bar shift=5pt,bar width=5.0pt,draw=teal!70!black,fill=teal!35] coordinates {(1,52.06) (2,178.70) (3,128.86) (4,224.18) (5,151.76) (6,292.35) (7,709.74) (8,443.64) (9,192.77) (10,542.05)};\addlegendentry{TCN}
\nextgroupplot[title={(b) Per-bearing coverage},ylabel={Coverage},ymin=0,ymax=1.05,xtick={1,2,3,4,5,6,7,8,9,10},xticklabels={B01,B02,B03,B04,B05,B08,B10,B11,B12,B17},xticklabel style={font=\scriptsize}]
\addplot[draw=none,fill=green!18] coordinates {(0.5,0.85) (10.5,0.85) (10.5,0.95) (0.5,0.95)} -- cycle;
\addplot+[mark=none,dashed,draw=gray!70] coordinates {(0.5,0.90) (10.5,0.90)};
\addplot+[only marks,mark=*,mark size=2.7pt,draw=blue!70!black,fill=blue!70!black] coordinates {(0.84,0.912) (1.84,0.791) (2.84,0.629) (3.84,0.947) (4.84,0.727) (5.84,0.938) (6.84,0.878) (7.84,0.860) (8.84,0.820) (9.84,0.708)};
\addplot+[only marks,mark=*,mark size=2.7pt,draw=purple!70!black,fill=purple!70!black] coordinates {(1.0,0.907) (2.0,0.898) (3.0,0.782) (4.0,0.905) (5.0,0.649) (6.0,0.943) (7.0,0.952) (8.0,0.881) (9.0,0.630) (10.0,0.765)};
\addplot+[only marks,mark=*,mark size=2.7pt,draw=teal!70!black,fill=teal!70!black] coordinates {(1.16,0.902) (2.16,0.965) (3.16,0.780) (4.16,0.678) (5.16,0.549) (6.16,0.839) (7.16,0.661) (8.16,0.608) (9.16,0.519) (10.16,0.671)};
\end{groupplot}
\end{tikzpicture}

\vspace{0.25em}
{\footnotesize Coverage markers are intentionally unconnected because bearing identifiers are categorical. The shaded band marks the primary 0.85--0.95 target range, which the predictive representation leave-bearing-out check does not satisfy on average.}
\caption{Per-bearing 10-bearing leave-bearing-out diagnostic performance.}
\label{fig:bearing_identity_diagnostic}
\end{figure}

Table~\ref{tab:bearing_identity_diagnostic} and Figure~\ref{fig:bearing_identity_diagnostic} report the 10-bearing leave-bearing-out generalization check. This evidence is deliberately separated from the primary leave-regime-out endpoint because it tests a different question: whether the method remains competitive when the held-out unit is bearing identity rather than operating regime. The calibrated predictive-representation model has the best mean MAE at 280.75, but the margin is small. Gradient-boosted tabular reaches MAE 284.98 and has better RMSE and interval score, while the random-subspace tabular baseline reaches MAE 289.18. Predictive-representation leave-bearing-out coverage is 0.821, below the predefined primary target band, and paired bearing-level tests are not significant against any baseline. The correct interpretation is supporting rather than conclusive bearing-disjoint evidence.

\begin{table}[!htbp]
\centering
\caption{10-bearing leave-bearing-out bearing-identity generalization check.}
\label{tab:bearing_identity_diagnostic}
\resizebox{\textwidth}{!}{%
\begin{tabular}{lrrrrrr}
\toprule
Model & Bearings & MAE & RMSE & Coverage & Width & Interval score \\
\midrule
Calibrated predictive representation & 10 & 280.75 & 350.31 & 0.821 & 946.25 & 1451.08 \\
Gradient-boosted tabular & 10 & 284.98 & 342.42 & 0.831 & 1010.19 & 1364.07 \\
Random-subspace ridge & 10 & 289.18 & 425.68 & 0.832 & 1026.66 & 1362.06 \\
TCN baseline & 10 & 291.61 & 365.07 & 0.717 & 745.00 & 1829.44 \\
Attention-and-weak-regularization baseline & 10 & 317.47 & 385.75 & 0.919 & 1231.56 & 1530.54 \\
\bottomrule
\end{tabular}
}

\vspace{0.25em}
{\footnotesize Leave-bearing-out is reported as a bearing-identity generalization check, not as the primary operating-regime shift endpoint.}
\end{table}

\FloatBarrier

\subsection{Matched-Protocol Context and Ablation Diagnostics}

The primary 10-bearing evaluation uses measured load/speed context and withholds derived regime identity from the model input. This input policy is methodological: derived regimes define reporting and held-out cells, not model features.

Table~\ref{tab:ablations} summarizes the matched-protocol diagnostic ablations. Removing load/speed context causes the largest degradation, increasing MAE from the matched diagnostic reference value of 273.89 to 405.78. This supports the practical value of measured operating metadata in the evaluated time-varying setting. Removing engineered features increases MAE to 292.24, indicating that conventional vibration descriptors still carry useful information even when raw signal encoding is available. Removing raw vibration increases MAE to 281.11, showing a smaller but still visible contribution from the raw signal stream. The interval-removal row is interpreted as an empirical-interval ablation from the same point model: its MAE is unchanged by construction, while coverage, width, and interval score are unavailable.

\begin{table}[!htbp]
\centering
\caption{Matched-protocol diagnostic ablation matrix for context, raw signal, engineered features, and empirical intervals.}
\label{tab:ablations}
\resizebox{\textwidth}{!}{%
\begin{tabular}{lrrrrrr}
\toprule
Variant & Regimes & MAE & RMSE & Coverage & Interval score & Mono. violations \\
\midrule
Reference 10-bearing configuration & 9 & 273.89 & 362.52 & 0.897 & 1859.90 & 0.494 \\
No load/speed context & 9 & 405.78 & 521.18 & 0.853 & 1825.00 & 0.495 \\
No raw vibration encoder & 9 & 281.11 & 368.82 & 0.902 & 1850.24 & 0.498 \\
No engineered features & 9 & 292.24 & 380.54 & 0.892 & 1925.83 & 0.496 \\
No empirical interval (same point model) & 9 & 273.89 & 362.52 & -- & -- & 0.494 \\
\bottomrule
\end{tabular}
}

\vspace{0.25em}
{\footnotesize These diagnostics use the matched model-selection/calibration protocol rather than the strict primary endpoint. The no-interval row removes intervals from the same point model, so its point MAE is unchanged and interval metrics are intentionally unavailable.}
\end{table}

\subsection{Matched-Protocol Prefix and Stress Diagnostics}

Table~\ref{tab:partial_robustness}, Figure~\ref{fig:partial_robustness}, and Figure~\ref{fig:coverage_profile} report 10-bearing prefix-observation, late-stage, and stress-test behavior. Prefix-observation results are mixed rather than monotone: MAE is 268.51 for windows in the first 20\% of life, 267.57 for the first 40\%, 254.13 for the first 60\%, and 285.78 for the first 80\%. Coverage remains inside the target band at 20\%, 60\%, and 80\%, and is slightly high at 40\% with empirical coverage 0.936. Late-stage windows have lower MAE at 207.53 and coverage 0.883. The non-monotone MAE pattern shows that prefix-observation difficulty depends on which regimes and bearings contribute to each life-fraction subset; Table~\ref{tab:prefix_composition} reports the subset composition behind those averages.

\begin{table}[!htbp]
\centering
\caption{Matched-protocol prefix-observation, noise, raw-channel-loss, and condition-response diagnostics.}
\label{tab:partial_robustness}
\resizebox{\textwidth}{!}{%
\begin{tabular}{lrrrrrrr}
\toprule
Setting & Regimes & MAE & RMSE & Coverage & Norm. width & Interval score & Width \\
\midrule
Matched diagnostic reference & 9 & 273.89 & 362.52 & 0.897 & 1.48 & 1859.90 & 1308.82 \\
Prefix observation: first 20\% & 9 & 268.51 & 321.27 & 0.908 & 0.79 & 1367.02 & 1176.51 \\
Prefix observation: first 40\% & 9 & 267.57 & 331.55 & 0.936 & 0.90 & 1459.44 & 1328.91 \\
Prefix observation: first 60\% & 9 & 254.13 & 325.13 & 0.920 & 1.03 & 1594.91 & 1354.65 \\
Prefix observation: first 80\% & 9 & 285.78 & 377.01 & 0.902 & 1.22 & 1899.52 & 1357.45 \\
Noise 5\% & 9 & 274.14 & 362.64 & 0.897 & 1.48 & 1857.42 & 1310.55 \\
Noise 10\% & 9 & 276.98 & 365.56 & 0.895 & 1.48 & 1854.32 & 1311.63 \\
Raw-channel loss & 9 & 421.01 & 542.65 & 0.741 & 1.64 & 3222.28 & 1439.85 \\
High load/speed response & 9 & 276.71 & 365.34 & -- & -- & -- & -- \\
Late-stage windows: life fraction $\geq$ 80\% & 9 & 207.53 & 274.86 & 0.883 & 5.51 & 1230.59 & 1016.06 \\
\bottomrule
\end{tabular}
}

\vspace{0.25em}
{\footnotesize These diagnostics use the matched model-selection/calibration protocol rather than the strict primary endpoint. Noise stress remains mild, while raw-channel-loss stress causes the largest degradation among the tested diagnostics and falls below the target coverage band. High load/speed response is a point-prediction sensitivity diagnostic; interval metrics are not computed.}
\end{table}

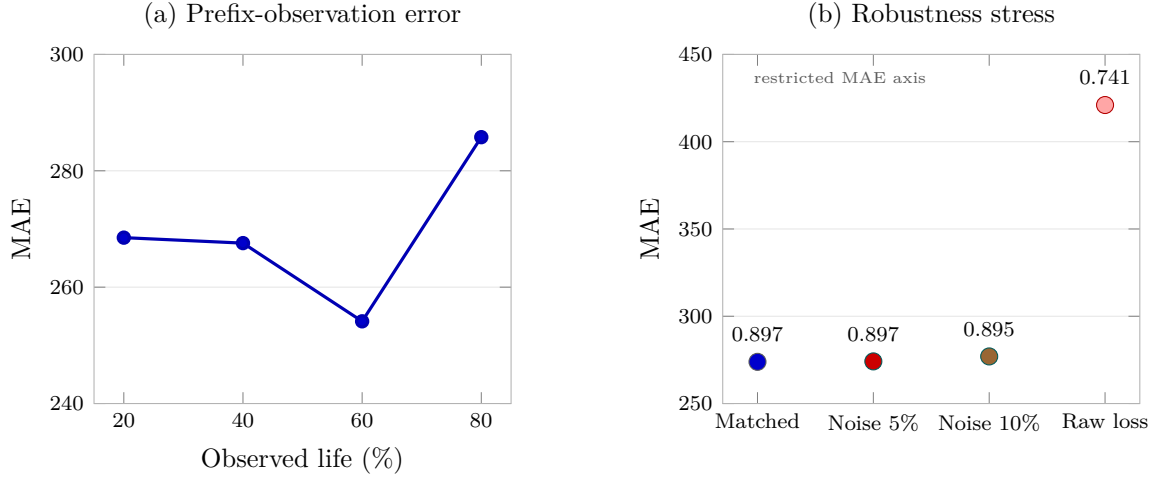
\begin{table}[!htbp]
\centering
\caption{Composition of 10-bearing prefix-observation subsets.}
\label{tab:prefix_composition}
\small
\begin{tabularx}{\textwidth}{lrrXr}
\toprule
Prefix subset & Windows & Bearings & Dominant regimes & Mean true RUL \\
\midrule
First 20\% & 2,864 & 10 & low-load / low-speed (642), low-load / high-speed (585) & 1819.8 \\
First 40\% & 5,721 & 10 & low-load / high-speed (1156), low-load / low-speed (946) & 1618.3 \\
First 60\% & 8,577 & 10 & low-load / high-speed (1728), mid-load / mid-speed (1177) & 1416.4 \\
First 80\% & 11,434 & 10 & low-load / high-speed (2230), mid-load / mid-speed (1529) & 1214.2 \\
\bottomrule
\end{tabularx}

\vspace{0.25em}
{\footnotesize Dominant regimes are the two largest load-speed cells within each prefix subset. This table contextualizes the non-monotone prefix-observation MAE pattern.}
\end{table}

\begin{figure}[!htbp]
\centering
\begin{tikzpicture}
\begin{groupplot}[
group style={group size=2 by 1,horizontal sep=28mm},
width=0.43\textwidth,
height=62mm,
tick label style={font=\scriptsize},
label style={font=\small},
title style={font=\small},
ymajorgrids=true,
grid style={gray!20},
axis line style={gray!60},
]
\nextgroupplot[
title={(a) Prefix-observation error},
xlabel={Observed life (\%)},
ylabel={MAE},
xmin=15,xmax=85,
ymin=240,ymax=300,
xtick={20,40,60,80},
]
\addplot+[very thick,mark=*,mark size=2.1pt,draw=blue!70!black] coordinates {(20,268.51) (40,267.57) (60,254.13) (80,285.78)};
\nextgroupplot[
title={(b) Robustness stress},
ylabel={MAE},
ymin=250,ymax=450,
xtick={1,2,3,4},
xticklabels={Matched,Noise 5\%,Noise 10\%,Raw loss},
xticklabel style={font=\scriptsize},
]
\addplot+[only marks,mark=*,mark size=3.2pt,draw=gray!70!black,fill=gray!30] coordinates {(1,273.89)};
\addplot+[only marks,mark=*,mark size=3.2pt,draw=teal!70!black,fill=teal!35] coordinates {(2,274.14)};
\addplot+[only marks,mark=*,mark size=3.2pt,draw=teal!70!black,fill=teal!50] coordinates {(3,276.98)};
\addplot+[only marks,mark=*,mark size=3.2pt,draw=red!70!black,fill=red!35] coordinates {(4,421.01)};
\node[font=\scriptsize,anchor=south] at (axis cs:1,279.89) {0.897};
\node[font=\scriptsize,anchor=south] at (axis cs:2,280.14) {0.897};
\node[font=\scriptsize,anchor=south] at (axis cs:3,282.98) {0.895};
\node[font=\scriptsize,anchor=south] at (axis cs:4,427.01) {0.741};
\node[font=\tiny,anchor=north west,text=gray!70!black] at (rel axis cs:0.05,0.98) {restricted MAE axis};
\end{groupplot}
\end{tikzpicture}

\vspace{0.25em}
{\footnotesize Numbers above the robustness markers are empirical interval coverage values. Noise 5\% and Noise 10\% denote 0.05 and 0.10 additive normalized signal noise; raw-channel loss denotes raw-stream ablation with engineered features retained. Panel (b) uses a restricted MAE axis for readability and is drawn as a dot plot rather than a zero-baseline bar chart. Raw-channel-loss stress causes the largest degradation among the tested diagnostics.}
\caption{Matched-protocol diagnostic prefix-observation and robustness MAE profile for the calibrated predictive-representation model.}
\label{fig:partial_robustness}
\end{figure}

\begin{figure}[!htbp]
\centering
\begin{tikzpicture}
\begin{groupplot}[
group style={group size=1 by 2,vertical sep=26mm},
width=0.92\textwidth,
height=39mm,
tick label style={font=\scriptsize},
label style={font=\small},
title style={font=\small,yshift=1mm},
ymajorgrids=true,
grid style={gray!20},
axis line style={gray!60},
]
\nextgroupplot[
title={(a) Coverage},
ymin=0.65,ymax=0.98,
ylabel={Empirical coverage},
xtick={1,2,3,4,5,6,7,8},
xticklabels={Matched,20\%,40\%,60\%,80\%,Noise 5\%,Noise 10\%,Raw loss},
xticklabel style={font=\scriptsize},
]
\addplot[draw=none,fill=green!12] coordinates {(0.5,0.85) (8.5,0.85) (8.5,0.95) (0.5,0.95)} -- cycle;
\addplot+[very thick,mark=*,mark size=3pt,draw=blue!70!black] coordinates {(1,0.897) (2,0.908) (3,0.936) (4,0.920) (5,0.902) (6,0.897) (7,0.895) (8,0.741)};
\node[font=\scriptsize,anchor=south east,fill=white,inner sep=1pt,text=red!70!black] at (axis cs:8,0.741) {0.741};
\addplot+[dashed,draw=gray!70] coordinates {(0.5,0.90) (8.5,0.90)};
\nextgroupplot[
title={(b) Interval score},
ylabel={Interval score},
ybar,
ymin=0,
xtick={1,2,3,4,5,6,7,8},
xticklabels={Matched,20\%,40\%,60\%,80\%,Noise 5\%,Noise 10\%,Raw loss},
xticklabel style={font=\scriptsize},
]
\addplot+[ybar,bar width=8pt,draw=teal!70!black,fill=teal!35] coordinates {(1,1860) (2,1367) (3,1459) (4,1595) (5,1900) (6,1857) (7,1854) (8,3222)};
\end{groupplot}
\end{tikzpicture}

\vspace{0.25em}
{\footnotesize The shaded band is the predefined 0.85--0.95 target range and the dashed line marks nominal 0.90 coverage. Noise 5\% and Noise 10\% denote 0.05 and 0.10 additive normalized signal noise; raw-channel loss denotes raw-stream ablation with engineered features retained. Raw-channel-loss stress fails both coverage and interval-score criteria.}
\caption{Matched-protocol diagnostic empirical-interval coverage and interval score across prefix-observation and robustness settings.}
\label{fig:coverage_profile}
\end{figure}

\FloatBarrier

Additive noise has limited effect in the matched-protocol 10-bearing stress suite: MAE changes from 273.89 in the matched diagnostic reference to 274.14 and 276.98 under 0.05 and 0.10 noise, with coverage remaining inside the target band. Raw-channel-loss stress is much more severe. MAE rises to 421.01 and coverage drops to 0.741, well below the target band. This is the largest degradation among the tested diagnostics: the framework does not establish reliable operation when the raw vibration stream is unavailable, even though engineered features and load/speed context are retained in this diagnostic.

\FloatBarrier

\subsection{Bearing-Regime Decision Risk Diagnostic}

Table~\ref{tab:bearing_level_decision} shifts the maintenance-risk diagnostic from pooled windows to bearing-regime groups. This is still retrospective, but it is a more reliability-relevant unit than individual windows because it counts whether each held-out bearing trajectory segment receives an on-time trigger before entering the last 20\% normalized RUL. For the calibrated predictive-representation model, a point-estimate trigger misses 18 of 43 late groups, while the lower-interval-bound trigger misses 3 of 43 late groups at the cost of 6 false-early groups. For the 400-tree random forest, the lower-bound trigger has no missed late groups but 8 false-early groups. Figure~\ref{fig:bearing_level_decision_risk} plots the corresponding risk over miss-to-early-inspection cost ratios $\rho\in\{1,2,5,10,20\}$. In this stylized bearing-regime risk diagnostic, the random-forest lower-bound trigger has lower risk than the predictive-representation lower-bound trigger at all tested miss-to-early-inspection cost ratios. The interval benefit for the predictive-representation model is therefore primarily within-model: the lower-bound trigger reduces missed-late groups relative to its own point-estimate trigger. The diagnostic does not establish a prospective alarm policy.

\begin{table}[!htbp]
\centering
\caption{Bearing-regime-level inspection-trigger diagnostic under the strict calibration split.}
\label{tab:bearing_level_decision}
\resizebox{\textwidth}{!}{%
\begin{tabular}{llrrrrrrrrrrr}
\toprule
Model & Trigger & Groups & Late groups & Missed late & False early & Median lead & Risk 1 & Risk 2 & Risk 5 & Risk 10 & Risk 20 \\
\midrule
Calibrated predictive representation & Point estimate & 53 & 43 & 18 & 0 & 0.176 & 33.96 & 67.92 & 169.81 & 339.62 & 679.25 \\
Calibrated predictive representation & Lower interval bound & 53 & 43 & 3 & 6 & 0.638 & 16.98 & 22.64 & 39.62 & 67.92 & 124.53 \\
400-tree random forest & Point estimate & 53 & 43 & 23 & 2 & 0.137 & 47.17 & 90.57 & 220.75 & 437.74 & 871.70 \\
400-tree random forest & Lower interval bound & 53 & 43 & 0 & 8 & 0.646 & 15.09 & 15.09 & 15.09 & 15.09 & 15.09 \\
\bottomrule
\end{tabular}
}

\vspace{0.25em}
{\footnotesize Rows aggregate over bearing-regime test groups rather than individual windows. A group is late if any saved test window has true normalized RUL at most $\tau=0.20$. Missed late counts groups without an on-time trigger at or before the first late window; false early counts groups with no late test window but at least one trigger. Risk values are per 100 bearing-regime groups, $L_\rho=100(\rho m+f)/n$, with missed-late count $m$, false-early count $f$, and miss-to-early-inspection cost ratio $\rho\in\{1,2,5,10,20\}$. This is a retrospective reliability-risk diagnostic, not a prospective maintenance policy.}
\end{table}

\begin{figure}[!htbp]
\centering
\begin{tikzpicture}
\begin{axis}[
width=0.86\textwidth,
height=0.42\textwidth,
xlabel={Miss-to-early-inspection cost ratio $\rho$},
ylabel={Risk per 100 bearing-regime groups},
xmin=0.5,xmax=20.5,
ymin=0,ymax=880,
xtick={1,2,5,10,20},
grid=both,
legend style={font=\scriptsize,at={(0.02,0.98)},anchor=north west,draw=none,fill=white,fill opacity=0.88,text opacity=1},
]
\addplot+[thick,mark=*,blue!70!black] coordinates {(1,33.96) (2,67.92) (5,169.81) (10,339.62) (20,679.25)};
\addplot+[thick,mark=square*,teal!70!black] coordinates {(1,16.98) (2,22.64) (5,39.62) (10,67.92) (20,124.53)};
\addplot+[thick,mark=triangle*,orange!80!black] coordinates {(1,47.17) (2,90.57) (5,220.75) (10,437.74) (20,871.70)};
\addplot+[thick,mark=diamond*,red!70!black] coordinates {(1,15.09) (2,15.09) (5,15.09) (10,15.09) (20,15.09)};
\legend{PR point,PR lower bound,RF point,RF lower bound}
\end{axis}
\end{tikzpicture}

\vspace{0.25em}
{\footnotesize Risk is computed from the same strict-split bearing-regime groups as Table~\ref{tab:bearing_level_decision}. PR denotes the calibrated predictive-representation model and RF denotes the 400-tree random forest. The curve is diagnostic and retrospective; it is not a prospective alarm-policy evaluation.}
\caption{Bearing-regime maintenance-risk curve under miss-to-early-inspection cost ratios.}
\label{fig:bearing_level_decision_risk}
\end{figure}

\FloatBarrier

\subsection{What the Evidence Supports and What It Does Not}

The results support a coherent but bounded conclusion. Under the strict split matrix with separate validation and calibration, the calibrated model attains the lowest mean normalized MAE while staying in the target coverage band, even though interval score is not uniformly best. Under matched selection and calibration, it remains the strongest neural accuracy--coverage compromise, with positive bootstrap relative-improvement intervals against the neural baselines and the less competitive tabular rows; exact nine-regime tests are not uniformly significant at 0.05, and the random forest still has lower mean MAE in the matched table. Load/speed context is essential in the ablations, and withholding derived regime identity is the more defensible primary input policy. Conditional, leave-bearing-out, and maintenance-risk diagnostics keep the model competitive while making concentration, undercoverage, and trigger tradeoffs visible.

The same evidence rules out stronger readings. It does not establish full 17-bearing PHME performance, XJTU-SY transfer, or NASA validation. It does not show that physics regularization drives accuracy, that the learned representation is a monotone physical damage variable, or that the method remains calibrated under raw-channel-loss sensor failure.

\section{Discussion}

\subsection{Reliability Evaluation Rather Than Architecture Novelty}

The main value of the study is the evaluation design, not an isolated network block. Regime-disjoint testing, residual calibration, paired regime statistics, context controls, fairness checks, conditional coverage, and maintenance-risk views are read together. Under this interpretation, the calibrated model is a competitive neural reference with target-band coverage on the strict endpoint, while the random forest remains close enough to prevent any claim of general point-error dominance. The matched protocol shows the same pattern: target-band neural coverage can be preferable even when a strong tabular model has lower raw MAE.

Failures are part of the evaluation, not secondary caveats. Raw-channel-loss and prefix-observation diagnostics mark where the model is usable and where it is fragile. Reporting the raw-channel-loss collapse keeps the primary coverage result from being mistaken for a sensor-fault guarantee.

\subsection{Why Physics Regularization Is Not the Mechanism Claim}

The experiments do not support a claim that physics regularization improves accuracy. Weakly regularized variants are competitive and somewhat more interpretable through their first-passage structure, but ablations do not isolate regularization as the driver. Monotonicity diagnostics remain weak, with health-score violation rates near 0.49. A stronger physics claim would need a structurally monotone architecture or a fatigue model tied more tightly to load, speed, cycles, and geometry. We therefore keep weak regularization as an optional inductive bias and as a diagnostic variant, not as the paper's mechanism claim.

\subsection{Reading Empirical Coverage Honestly}

The strict endpoint reaches 0.900 empirical coverage for nominal 90\% intervals. That average is useful, but it is not a distribution-free guarantee under arbitrary shift, and it is not uniform. Coverage falls to 0.666 in low-load/high-speed, often with high single-bearing concentration. The pooled regime-conditioned residual diagnostic repairs that cell only as a post-hoc cross-rotation probe. Raw-channel-loss stress drops coverage to 0.741. Prefix coverage is mostly inside the target band, but MAE is non-monotone across life fractions. Average calibration is therefore necessary but not sufficient for reliability claims.

\subsection{Maintenance Triggers Without Policy Overreach}

Intervals matter because they change trigger behavior, not because they automatically define a safe policy. Table~\ref{tab:bearing_level_decision} converts the strict-split intervals into a retrospective bearing-regime risk view. For the calibrated model, switching from a point trigger to a lower-bound trigger reduces missed late groups from 18 to 3 at the cost of 6 false-early groups. The random-forest lower-bound trigger misses none of the late groups but produces 8 false-early groups and has lower stylized risk at all tested miss-to-early cost ratios. The interval benefit for the predictive-representation model is therefore mainly within-model: the lower bound improves on its own point trigger. No safety-critical maintenance policy is claimed.

\subsection{Keeping the Evidence Boundary Visible}

The study stops at the processed 10-bearing subset by design. That boundary is a feature of the evaluation, not an accidental omission. Full-archive PHME processing is the natural next step if the goal becomes broad benchmark coverage; external sets such as XJTU-SY remain secondary until the primary time-varying evidence is strengthened.

\section{Limitations}

The evidence stops at ten PHME runs---B01, B02, B03, B04, B05, B08, B10, B11, B12, and B17. That is the intended scope, not a proxy for the full 17-bearing public record. Seven public bearings remain unprocessed, so the subset cannot be proven unbiased. No XJTU-SY, IMS, FEMTO, or C-MAPSS validation is included, and full-dataset or cross-dataset claims would be premature.

The primary split tests operating-regime shift, not every form of generalization. Leave-bearing-out is only a supporting identity check, and paired bearing-level tests are not significant against the strongest tabular baselines. The model is also not a validated physical damage-state estimator: weak monotonicity and smoothness terms are heuristic biases, not a complete fatigue model, and absolute-step MAE is retrospective because normalized predictions are scaled by each bearing's known maximum step count for offline reporting.

Calibration is empirical and incomplete. Average strict coverage reaches the target band, but low-load/high-speed coverage falls to 0.666 and raw-channel-loss coverage falls to 0.741. The pooled regime-conditioned residual diagnostic improves conditional coverage only post hoc and is not a primary interval or a formal single-split Mondrian guarantee. Stress, seed, and maintenance diagnostics are likewise bounded: they probe selected faults and retrospective triggers rather than all sensor failures or prospective alarm policies. The fixed one-channel, 512-sample preprocessing design should be read as the study design, not as an optimized architectural conclusion.

\section{Conclusion}

This paper asked a reliability question about bearing RUL under operating-regime shift: when load and speed change, do both point accuracy and prediction intervals remain trustworthy? On a documented 10-bearing PHME subset, a calibrated predictive-representation model provides a competitive answer under a strict leave-operating-regime-out protocol. It attains the lowest mean normalized MAE while preserving target coverage on the separate-validation/separate-calibration matrix, with only a modest margin over random forest and standard gradient-boosted regression. Under matched selection and calibration, it remains the strongest neural accuracy--coverage compromise in this diagnostic, while the 400-tree random forest achieves lower MAE but falls below the coverage target. Conditional undercoverage and raw-channel-loss failure are reported as central evidence, not secondary caveats. These conclusions apply only to the processed ten-bearing subset; they do not establish full-archive PHME performance, cross-dataset transfer, or deployment-ready maintenance policy.

The contribution is therefore a bounded reliability-evaluation protocol with explicit failure modes, not a new digital-twin theory or a full-PHME state-of-the-art claim. Open work remains clear: process the remaining public PHME bearings, move from post-hoc to pre-specified conditional calibration, strengthen sensor-fault probes, widen seed replication, and only then seek external validation and prospective maintenance-policy evaluation.

\section*{Declarations}

\noindent\textbf{Funding:} This research is supported by Santa Clara University.

\noindent\textbf{Declaration of competing interests:} The authors declare no competing interests.

\noindent\textbf{Author contributions:} Shaoliang Yang: Conceptualization, Methodology, Software, Data curation, Formal analysis, Investigation, Visualization, Writing - original draft. Jun Wang: Conceptualization, Supervision, Project administration, Resources, Methodology, Writing - review and editing. Yunsheng Wang: Validation, Investigation, Methodology, Writing - review and editing.

\noindent\textbf{Data availability statement:} The original PHME bearing data are available from the public dataset source cited in the manuscript. The full public code release for this study is available at \publiccodereleaseurl. The repository contains the implementation, experiment configurations, tests, analysis utilities, and reproduction instructions needed to rerun the reported analyses. The public release does not redistribute the PHME data or derived prediction and model-output files, which are governed by dataset licensing and venue-sharing constraints. Additional numerical outputs used to verify the reported tables and figures are available from the authors where those constraints permit.

\noindent\textbf{Acknowledgements:} None.

\bibliographystyle{plainnat}
\bibliography{references}

@article{lei2018machinery_review,
  author = {Lei, Yaguo and Li, Naipeng and Guo, Liang and Li, Ningbo and Yan, Tao and Lin, Jun},
  title = {Machinery Health Prognostics: A Systematic Review from Data Acquisition to {RUL} Prediction},
  journal = {Mechanical Systems and Signal Processing},
  volume = {104},
  pages = {799--834},
  year = {2018},
  doi = {10.1016/j.ymssp.2017.11.016}
}

@article{wang2020hybrid_bearing,
  author = {Wang, Biao and Lei, Yaguo and Li, Naipeng and Li, Ningbo},
  title = {A Hybrid Prognostics Approach for Estimating Remaining Useful Life of Rolling Element Bearings},
  journal = {IEEE Transactions on Reliability},
  volume = {69},
  number = {1},
  pages = {401--412},
  year = {2020},
  doi = {10.1109/TR.2018.2882682}
}

@article{li2019multiscale_bearing,
  author = {Li, Xiang and Zhang, Wei and Ding, Qian},
  title = {Deep Learning-Based Remaining Useful Life Estimation of Bearings Using Multi-Scale Feature Extraction},
  journal = {Reliability Engineering \& System Safety},
  volume = {182},
  pages = {208--218},
  year = {2019},
  doi = {10.1016/j.ress.2018.11.011}
}

@article{li2019twofactor_tvoc,
  author = {Li, Naipeng and Gebraeel, Nagi and Lei, Yaguo and Bian, Linkan and Si, Xiaosheng},
  title = {Remaining Useful Life Prediction of Machinery under Time-Varying Operating Conditions Based on a Two-Factor State-Space Model},
  journal = {Reliability Engineering \& System Safety},
  volume = {186},
  pages = {88--100},
  year = {2019},
  doi = {10.1016/j.ress.2019.02.017}
}

@article{dacosta2020domain_adaptation,
  author = {da Costa, Paulo Roberto de Oliveira and Ak{\c{c}}ay, Alp and Zhang, Yingqian and Kaymak, Uzay},
  title = {Remaining Useful Lifetime Prediction via Deep Domain Adaptation},
  journal = {Reliability Engineering \& System Safety},
  volume = {195},
  pages = {106682},
  year = {2020},
  doi = {10.1016/j.ress.2019.106682}
}

@article{zhang2021representation_regularization,
  author = {Zhang, Wei and Li, Xiang and Ma, Hui and Luo, Zhong and Li, Xu},
  title = {Transfer Learning Using Deep Representation Regularization in Remaining Useful Life Prediction across Operating Conditions},
  journal = {Reliability Engineering \& System Safety},
  volume = {211},
  pages = {107556},
  year = {2021},
  doi = {10.1016/j.ress.2021.107556}
}

@article{ding2021deep_metric_transfer,
  author = {Ding, Yifei and Jia, Minping and Miao, Qiuhua and Huang, Peng},
  title = {Remaining Useful Life Estimation Using Deep Metric Transfer Learning for Kernel Regression},
  journal = {Reliability Engineering \& System Safety},
  volume = {212},
  pages = {107583},
  year = {2021},
  doi = {10.1016/j.ress.2021.107583}
}

@article{javanmardi2024phme_tvoc,
  author = {Javanmardi, Alireza and Aimiyekagbon, Osarenren Kennedy and Bender, Amelie and Kimotho, James Kuria and Sextro, Walter and H{\"u}llermeier, Eyke},
  title = {Remaining Useful Lifetime Estimation of Bearings Operating under Time-Varying Conditions},
  journal = {PHM Society European Conference},
  volume = {8},
  number = {1},
  pages = {9},
  year = {2024},
  doi = {10.36001/phme.2024.v8i1.4101}
}

@misc{aimiyekagbon2024zenodo,
  author = {Aimiyekagbon, Osarenren Kennedy},
  title = {Run-to-Failure Data Set of Ball Bearings Subjected to Time-Varying Load and Speed Conditions},
  howpublished = {Zenodo dataset},
  year = {2024},
  doi = {10.5281/zenodo.10868257},
  url = {https://zenodo.org/records/10868257}
}

@article{breiman2001random_forests,
  author = {Breiman, Leo},
  title = {Random Forests},
  journal = {Machine Learning},
  volume = {45},
  number = {1},
  pages = {5--32},
  year = {2001},
  doi = {10.1023/A:1010933404324}
}

@article{friedman2001greedy,
  author = {Friedman, Jerome H.},
  title = {Greedy Function Approximation: A Gradient Boosting Machine},
  journal = {The Annals of Statistics},
  volume = {29},
  number = {5},
  pages = {1189--1232},
  year = {2001},
  doi = {10.1214/aos/1013203451}
}

@article{bai2018tcn,
  author = {Bai, Shaojie and Kolter, J. Zico and Koltun, Vladlen},
  title = {An Empirical Evaluation of Generic Convolutional and Recurrent Networks for Sequence Modeling},
  journal = {arXiv preprint arXiv:1803.01271},
  year = {2018},
  doi = {10.48550/arXiv.1803.01271},
  url = {https://arxiv.org/abs/1803.01271}
}

@article{pedregosa2011scikit,
  author = {Pedregosa, Fabian and Varoquaux, Ga{\"e}l and Gramfort, Alexandre and Michel, Vincent and Thirion, Bertrand and Grisel, Olivier and Blondel, Mathieu and Prettenhofer, Peter and Weiss, Ron and Dubourg, Vincent and Vanderplas, Jake and Passos, Alexandre and Cournapeau, David and Brucher, Matthieu and Perrot, Matthieu and Duchesnay, {\'E}douard},
  title = {Scikit-Learn: Machine Learning in Python},
  journal = {Journal of Machine Learning Research},
  volume = {12},
  pages = {2825--2830},
  year = {2011},
  url = {https://jmlr.org/papers/v12/pedregosa11a.html}
}

@article{cui2024dt_graph_da,
  author = {Cui, Lingli and Xiao, Yongchang and Liu, Dongdong and Han, Honggui},
  title = {Digital Twin-Driven Graph Domain Adaptation Neural Network for Remaining Useful Life Prediction of Rolling Bearing},
  journal = {Reliability Engineering \& System Safety},
  volume = {245},
  pages = {109991},
  year = {2024},
  doi = {10.1016/j.ress.2024.109991}
}

@article{hou2024cross_transformer_segmented_cleaning,
  author = {Hou, Dongxiao and Chen, JiaHui and Cheng, Rongcai and Hu, Xue and Shi, Peiming},
  title = {A Bearing Remaining Life Prediction Method under Variable Operating Conditions Based on Cross-Transformer Fusioning Segmented Data Cleaning},
  journal = {Reliability Engineering \& System Safety},
  volume = {245},
  pages = {110021},
  year = {2024},
  doi = {10.1016/j.ress.2024.110021}
}

@article{wang2024mtl_mdn_bearing_interval,
  author = {Wang, Xin and Li, Yongbo and Noman, Khandaker and Nandi, Asoke K.},
  title = {Multi-Task Learning Mixture Density Network for Interval Estimation of the Remaining Useful Life of Rolling Element Bearings},
  journal = {Reliability Engineering \& System Safety},
  volume = {251},
  pages = {110348},
  year = {2024},
  doi = {10.1016/j.ress.2024.110348}
}

@article{zhan2024mdf_uq,
  author = {Zhan, Yuling and Kong, Ziqian and Wang, Ziqi and Jin, Xiaohang and Xu, Zhengguo},
  title = {Remaining Useful Life Prediction with Uncertainty Quantification Based on Multi-Distribution Fusion Structure},
  journal = {Reliability Engineering \& System Safety},
  volume = {251},
  pages = {110383},
  year = {2024},
  doi = {10.1016/j.ress.2024.110383}
}

@article{li2025reliable_bearing_mhde,
  author = {Li, Wenjie and Liu, Dongdong and Wang, Xin and Cui, Lingli},
  title = {A Reliable Bearing Remaining Useful Life Prediction Method Based on Multi-Hierarchy Dynamic Evaluation and Uncertainty Amelioration},
  journal = {Reliability Engineering \& System Safety},
  volume = {263},
  pages = {111270},
  year = {2025},
  doi = {10.1016/j.ress.2025.111270}
}

@article{gong2025dt_pinn_uq,
  author = {Gong, Fengjin and Ma, Ping and Zhang, Hongli and Wang, Cong and Li, Xinkai and Wu, Yinfei},
  title = {Rolling Bearings Remaining Useful Life Estimation Using Digital Twin and Physics-Informed Methods with Uncertainty Quantification},
  journal = {Engineering Applications of Artificial Intelligence},
  volume = {154},
  pages = {111070},
  year = {2025},
  doi = {10.1016/j.engappai.2025.111070}
}

@article{bott2026uncertainty_bearing_flows,
  author = {Bott, Alexander and Liu, Bolin and Nuding, Linus and Wachsmuth, Julian and Puchta, Alexander and Fleischer, J{\"u}rgen},
  title = {Uncertainty-Aware Prognostics of Ball Bearings Using Physics-Based Simulation and Conditional Normalizing Flows},
  journal = {IEEE Access},
  volume = {14},
  pages = {20100--20110},
  year = {2026},
  doi = {10.1109/ACCESS.2026.3661174}
}

@article{li2024pidd_rul_review,
  author = {Li, Huiqin and Zhang, Zhengxin and Li, Tianmei and Si, Xiaosheng},
  title = {A Review on Physics-Informed Data-Driven Remaining Useful Life Prediction: Challenges and Opportunities},
  journal = {Mechanical Systems and Signal Processing},
  volume = {209},
  pages = {111120},
  year = {2024},
  doi = {10.1016/j.ymssp.2024.111120}
}

@article{javanmardi2023conformal_rul,
  author = {Javanmardi, Alireza and H{\"u}llermeier, Eyke},
  title = {Conformal Prediction Intervals for Remaining Useful Lifetime Estimation},
  journal = {International Journal of Prognostics and Health Management},
  volume = {14},
  number = {2},
  year = {2023},
  doi = {10.36001/ijphm.2023.v14i2.3417}
}

@article{jardine2006cbm,
  author = {Jardine, Andrew K. S. and Lin, Daming and Banjevic, Dragan},
  title = {A Review on Machinery Diagnostics and Prognostics Implementing Condition-Based Maintenance},
  journal = {Mechanical Systems and Signal Processing},
  volume = {20},
  number = {7},
  pages = {1483--1510},
  year = {2006},
  doi = {10.1016/j.ymssp.2005.09.012}
}

@article{heng2009rotating,
  author = {Heng, Aiwina and Zhang, Sheng and Tan, Andy C. C. and Mathew, Joseph},
  title = {Rotating Machinery Prognostics: State of the Art, Challenges and Opportunities},
  journal = {Mechanical Systems and Signal Processing},
  volume = {23},
  number = {3},
  pages = {724--739},
  year = {2009},
  doi = {10.1016/j.ymssp.2008.06.009}
}

@article{sikorska2011prognostic,
  author = {Sikorska, Joanna Z. and Hodkiewicz, Melinda and Ma, Lin},
  title = {Prognostic Modelling Options for Remaining Useful Life Estimation by Industry},
  journal = {Mechanical Systems and Signal Processing},
  volume = {25},
  number = {5},
  pages = {1803--1836},
  year = {2011},
  doi = {10.1016/j.ymssp.2010.11.018}
}

@article{si2011statistical_rul,
  author = {Si, Xiao-Sheng and Wang, Wenbin and Hu, Chang-Hua and Zhou, Dong-Hua},
  title = {Remaining Useful Life Estimation: A Review on the Statistical Data Driven Approaches},
  journal = {European Journal of Operational Research},
  volume = {213},
  number = {1},
  pages = {1--14},
  year = {2011},
  doi = {10.1016/j.ejor.2010.11.018}
}

@article{zio2022phm,
  author = {Zio, Enrico},
  title = {Prognostics and Health Management ({PHM}): Where Are We and Where Do We (Need to) Go in Theory and Practice},
  journal = {Reliability Engineering \& System Safety},
  volume = {218},
  pages = {108119},
  year = {2022},
  doi = {10.1016/j.ress.2021.108119}
}

@article{hu2022phm_review,
  author = {Hu, Yang and Miao, Xuewen and Si, Yong and Pan, Ershun and Zio, Enrico},
  title = {Prognostics and Health Management: A Review from the Perspectives of Design, Development and Decision},
  journal = {Reliability Engineering \& System Safety},
  volume = {217},
  pages = {108063},
  year = {2022},
  doi = {10.1016/j.ress.2021.108063}
}

@article{sankararaman2015uncertainty,
  author = {Sankararaman, Shankar},
  title = {Significance, Interpretation, and Quantification of Uncertainty in Prognostics and Remaining Useful Life Prediction},
  journal = {Mechanical Systems and Signal Processing},
  volume = {52--53},
  pages = {228--247},
  year = {2015},
  doi = {10.1016/j.ymssp.2014.05.029}
}

@article{lei2018distribution_free,
  author = {Lei, Jing and G'Sell, Max and Rinaldo, Alessandro and Tibshirani, Ryan J. and Wasserman, Larry},
  title = {Distribution-Free Predictive Inference for Regression},
  journal = {Journal of the American Statistical Association},
  volume = {113},
  number = {523},
  pages = {1094--1111},
  year = {2018},
  doi = {10.1080/01621459.2017.1307116}
}

@inproceedings{romano2019cqr,
  author = {Romano, Yaniv and Patterson, Evan and Cand{\`e}s, Emmanuel},
  title = {Conformalized Quantile Regression},
  booktitle = {Advances in Neural Information Processing Systems},
  volume = {32},
  pages = {3538--3548},
  year = {2019}
}

@misc{angelopoulos2021gentle,
  author = {Angelopoulos, Anastasios N. and Bates, Stephen},
  title = {A Gentle Introduction to Conformal Prediction and Distribution-Free Uncertainty Quantification},
  year = {2021},
  eprint = {2107.07511},
  archivePrefix = {arXiv},
  primaryClass = {cs.LG},
  doi = {10.48550/arXiv.2107.07511},
  url = {https://arxiv.org/abs/2107.07511}
}

@article{zhao2019deep_health,
  author = {Zhao, Rui and Yan, Ruqiang and Chen, Zhenghua and Mao, Kezhi and Wang, Peng and Gao, Robert X.},
  title = {Deep Learning and Its Applications to Machine Health Monitoring},
  journal = {Mechanical Systems and Signal Processing},
  volume = {115},
  pages = {213--237},
  year = {2019},
  doi = {10.1016/j.ymssp.2018.05.050}
}

@article{fink2020deep_phm,
  author = {Fink, Olga and Wang, Qin and Svens{\'e}n, Markus and Dersin, Pierre and Lee, Wan-Jui and Ducoffe, Melanie},
  title = {Potential, Challenges and Future Directions for Deep Learning in Prognostics and Health Management Applications},
  journal = {Engineering Applications of Artificial Intelligence},
  volume = {92},
  pages = {103678},
  year = {2020},
  doi = {10.1016/j.engappai.2020.103678}
}

@article{li2018dcnn_rul,
  author = {Li, Xiang and Ding, Qian and Sun, Jian-Qiao},
  title = {Remaining Useful Life Estimation in Prognostics Using Deep Convolution Neural Networks},
  journal = {Reliability Engineering \& System Safety},
  volume = {172},
  pages = {1--11},
  year = {2018},
  doi = {10.1016/j.ress.2017.11.021}
}

@article{zhang2018lstm_rul,
  author = {Zhang, Jianjing and Wang, Peng and Yan, Ruqiang and Gao, Robert X.},
  title = {Long Short-Term Memory for Machine Remaining Life Prediction},
  journal = {Journal of Manufacturing Systems},
  volume = {48},
  pages = {78--86},
  year = {2018},
  doi = {10.1016/j.jmsy.2018.05.011}
}

@article{ren2018dcnn_bearing,
  author = {Ren, Lei and Sun, Yaqiang and Wang, Hao and Zhang, Lin},
  title = {Prediction of Bearing Remaining Useful Life with Deep Convolution Neural Network},
  journal = {IEEE Access},
  volume = {6},
  pages = {13041--13049},
  year = {2018},
  doi = {10.1109/ACCESS.2018.2804930}
}

@article{guo2017rnn_hi,
  author = {Guo, Liang and Li, Naipeng and Jia, Feng and Lei, Yaguo and Lin, Jing},
  title = {A Recurrent Neural Network Based Health Indicator for Remaining Useful Life Prediction of Bearings},
  journal = {Neurocomputing},
  volume = {240},
  pages = {98--109},
  year = {2017},
  doi = {10.1016/j.neucom.2017.02.045}
}

@article{wang2019dscn,
  author = {Wang, Biao and Lei, Yaguo and Li, Naipeng and Yan, Tao},
  title = {Deep Separable Convolutional Network for Remaining Useful Life Prediction of Machinery},
  journal = {Mechanical Systems and Signal Processing},
  volume = {134},
  pages = {106330},
  year = {2019},
  doi = {10.1016/j.ymssp.2019.106330}
}

@article{li2021hagcn,
  author = {Li, Tianfu and Zhao, Zhibin and Sun, Chuang and Yan, Ruqiang and Chen, Xuefeng},
  title = {Hierarchical Attention Graph Convolutional Network to Fuse Multi-Sensor Signals for Remaining Useful Life Prediction},
  journal = {Reliability Engineering \& System Safety},
  volume = {215},
  pages = {107878},
  year = {2021},
  doi = {10.1016/j.ress.2021.107878}
}

@article{li2022gnn_benchmark,
  author = {Li, Tianfu and Zhou, Zheng and Li, Sinan and Sun, Chuang and Yan, Ruqiang and Chen, Xuefeng},
  title = {The Emerging Graph Neural Networks for Intelligent Fault Diagnostics and Prognostics: A Guideline and a Benchmark Study},
  journal = {Mechanical Systems and Signal Processing},
  volume = {168},
  pages = {108653},
  year = {2022},
  doi = {10.1016/j.ymssp.2021.108653}
}

@inproceedings{nectoux2012pronostia,
  author = {Nectoux, Patrick and Gouriveau, Rafael and Medjaher, Kamal and Ramasso, Emmanuel and Chebel-Morello, Brigitte and Zerhouni, Noureddine and Varnier, Christophe},
  title = {{PRONOSTIA}: An Experimental Platform for Bearings Accelerated Degradation Tests},
  booktitle = {IEEE International Conference on Prognostics and Health Management},
  pages = {1--8},
  year = {2012}
}

@misc{nasa2025ims,
  author = {{NASA Prognostics Center of Excellence}},
  title = {{IMS Bearings} Dataset},
  howpublished = {NASA Open Data Portal},
  year = {2025},
  url = {https://data.nasa.gov/dataset/ims-bearings},
  note = {Accessed 30 April 2026}
}

@article{fuller2020digital_twin,
  author = {Fuller, Aidan and Fan, Zhong and Day, Charles and Barlow, Chris},
  title = {Digital Twin: Enabling Technologies, Challenges and Open Research},
  journal = {IEEE Access},
  volume = {8},
  pages = {108952--108971},
  year = {2020},
  doi = {10.1109/ACCESS.2020.2998358}
}

@article{tao2019digital_twin,
  author = {Tao, Fei and Zhang, He and Liu, Ang and Nee, Andrew Y. C.},
  title = {Digital Twin in Industry: State-of-the-Art},
  journal = {IEEE Transactions on Industrial Informatics},
  volume = {15},
  number = {4},
  pages = {2405--2415},
  year = {2019},
  doi = {10.1109/TII.2018.2873186}
}

@article{karniadakis2021piml,
  author = {Karniadakis, George Em and Kevrekidis, Ioannis G. and Lu, Lu and Perdikaris, Paris and Wang, Sifan and Yang, Liu},
  title = {Physics-Informed Machine Learning},
  journal = {Nature Reviews Physics},
  volume = {3},
  number = {6},
  pages = {422--440},
  year = {2021},
  doi = {10.1038/s42254-021-00314-5}
}

@article{raissi2019pinn,
  author = {Raissi, Maziar and Perdikaris, Paris and Karniadakis, George Em},
  title = {Physics-Informed Neural Networks: A Deep Learning Framework for Solving Forward and Inverse Problems Involving Nonlinear Partial Differential Equations},
  journal = {Journal of Computational Physics},
  volume = {378},
  pages = {686--707},
  year = {2019},
  doi = {10.1016/j.jcp.2018.10.045}
}

@article{liao2023attnpinn,
  author = {Liao, Xinyuan and Chen, Shaowei and Wen, Pengfei and Zhao, Shuai},
  title = {Remaining Useful Life with Self-Attention Assisted Physics-Informed Neural Network},
  journal = {Advanced Engineering Informatics},
  volume = {58},
  pages = {102195},
  year = {2023},
  doi = {10.1016/j.aei.2023.102195}
}

@article{he2023pignn,
  author = {He, Yuxuan and Su, Huai and Zio, Enrico and Peng, Shiliang and Fan, Lin and Yang, Zhaoming and Yang, Zhe and Zhang, Jinjun},
  title = {A Systematic Method of Remaining Useful Life Estimation Based on Physics-Informed Graph Neural Networks with Multisensor Data},
  journal = {Reliability Engineering \& System Safety},
  volume = {237},
  pages = {109333},
  year = {2023},
  doi = {10.1016/j.ress.2023.109333}
}

@article{beaulieu2024physics_aug,
  author = {de Beaulieu, Martin Herv{\'e} and Jha, Mayank Shekhar and Garnier, Hugues and Cerbah, Farid},
  title = {Remaining Useful Life Prediction Based on Physics-Informed Data Augmentation},
  journal = {Reliability Engineering \& System Safety},
  volume = {252},
  pages = {110451},
  year = {2024},
  doi = {10.1016/j.ress.2024.110451}
}

@article{lu2023dt_bearing,
  author = {Lu, Quanbo and Li, Mei},
  title = {Digital Twin-Driven Remaining Useful Life Prediction for Rolling Element Bearing},
  journal = {Machines},
  volume = {11},
  number = {7},
  pages = {678},
  year = {2023},
  doi = {10.3390/machines11070678}
}

\end{document}